\documentclass[%
 pof,
 amsmath,amssymb,
 reprint,%
]{revtex4-2}

\usepackage{graphicx}%
\usepackage{dcolumn}%
\usepackage{bm}%

\usepackage[utf8]{inputenc}
\usepackage[T1]{fontenc}
\usepackage{mathptmx}
\usepackage{etoolbox}

\usepackage{tikz}
\usetikzlibrary{external}
\usetikzlibrary{colorbrewer}
\usepackage{pgfplots}
\pgfplotsset{compat=1.16}

\usepgfplotslibrary{external} 
\usepgfplotslibrary{groupplots}
\usepgfplotslibrary{colorbrewer}
\usepgfplotslibrary{fillbetween}

\graphicspath{figures/} %

\usepackage{tabularx}   %
\usepackage{booktabs}   %
\usepackage{siunitx}    %
\usepackage{dsfont}     %

\usepackage{ifthen}
\newboolean{final}
\setboolean{final}{true}

\definecolor{Dark2-color0}{RGB}{  0,  0,  0}
\definecolor{Dark2-color1}{RGB}{ 27,158,119}
\definecolor{Dark2-color2}{RGB}{217, 95,  2}
\definecolor{Dark2-color3}{RGB}{117,112,179}
\definecolor{Dark2-color4}{RGB}{231, 41,138}
\definecolor{Dark2-color5}{RGB}{102,166, 30}
\definecolor{Dark2-color6}{RGB}{230,171,  2}

\pgfplotscreateplotcyclelist{DarkTwo}{%
  thick, Dark2-color0, solid, mark=none \\%
  thick, Dark2-color2, solid, mark=none \\%
  thick, Dark2-color1, solid, mark=none \\%
  thick, Dark2-color3, solid, mark=none \\%
  thick, Dark2-color4, solid, mark=none \\%
  thick, Dark2-color5, solid, mark=none \\%
  thick, Dark2-color6, solid, mark=none \\%
}

\usepackage[hidelinks]{hyperref}

\newcommand{\figref}[1]{Figure~\ref{#1}}
\newcommand{\tabref}[1]{Table~\ref{#1}}
\newcommand{\secref}[1]{Section~\ref{#1}}
\LetLtxMacro{\originaleqref}{\eqref}
\renewcommand{\eqref}{Eq.~\originaleqref}

\usepackage{xcolor}

\newcommand{\ppvec}[1]{\bm{#1}}

\makeatletter
\def\@email#1#2{%
 \endgroup
 \patchcmd{\titleblock@produce}
  {\frontmatter@RRAPformat}
  {\frontmatter@RRAPformat{\produce@RRAP{*#1\href{mailto:#2}{#2}}}\frontmatter@RRAPformat}
  {}{}
}%
\makeatother
\begin{document}

\preprint{AIP/123-QED}

\title[Closure Schemes for LES Using RL]{Toward Discretization-Consistent Closure Schemes for Large Eddy Simulation Using Reinforcement Learning}
\author{Andrea Beck}%
\author{Marius Kurz\footnote{Both authors cordially agree to share first authorship.}}

\email{\{beck,m.kurz\}@iag.uni-stuttgart.de}
\affiliation{ 
Institute of Aerodynamics and Gas Dynamics, University of Stuttgart, Pfaﬀenwaldring 21, 70569 Stuttgart, Germany
}%

\date{\today}%

\begin{abstract}
This study proposes a novel method for developing discretization-consistent closure schemes for implicitly filtered Large Eddy Simulation (LES).
Here, the induced filter kernel, and thus the closure terms, are determined by the properties of the grid and the discretization operator, leading to additional computational subgrid terms that are generally unknown in a priori analysis.
In this work, the task of adapting the coefficients of LES closure models is thus framed as a Markov decision process and solved in an a posteriori manner with Reinforcement Learning~(RL).
This optimization framework is applied to both explicit and implicit closure models.
The explicit model is based on an element-local eddy viscosity model.
The optimized model is found to adapt its induced viscosity within discontinuous Galerkin~(DG) methods to homogenize the dissipation within an element by adding more viscosity near its center.
  For the implicit modeling, RL is applied to identify an optimal blending strategy for a hybrid DG and Finite Volume~(FV) scheme.
The resulting optimized discretization yields more accurate results in LES than either the pure DG or FV method and renders itself as a viable modeling ansatz that could initiate a novel class of high-order schemes for compressible turbulence by combining turbulence modeling with shock capturing in a single framework.
All newly derived models achieve accurate results that either match or outperform traditional models for different discretizations and resolutions.
Overall, the results demonstrate that the proposed RL optimization can provide discretization-consistent closures that could reduce the uncertainty in implicitly filtered LES.
\end{abstract}

\maketitle

\section{Introduction}
\begin{figure*}
	\centering
  \ifthenelse{\boolean{final}}{
    \includegraphics[width=0.71\linewidth]{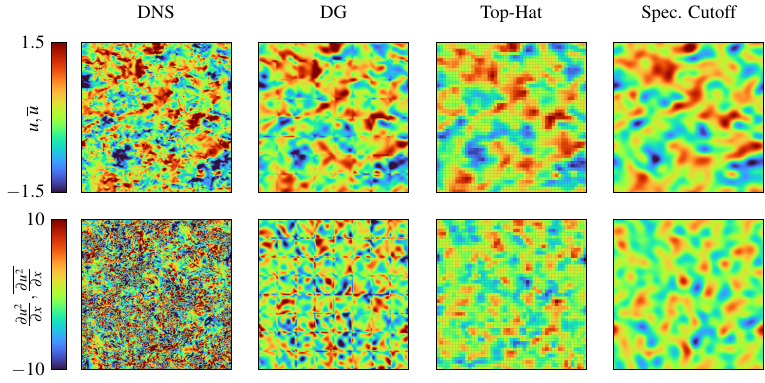}
  }{
    \tikzsetnextfilename{fig_tikz_filter}
    \input{tikz/fig_filter.tikz}
  }
  \caption{Slices of the instantaneous velocity field in $x$-direction $u$~(\emph{top left}) of a DNS flow field of Homogeneous Isotropic Turbulence~(HIT) and the corresponding component of the filtered nonlinear convective fluxes of the Navier--Stokes equations, which corresponds to the term \smash{$\frac{\partial u^2}{\partial x}$} in \eqref{eq:burgers}. The other three columns show the corresponding fields after applying (\emph{from left to right}) either an $L_2$-projection filter onto an elementwise polynomial basis (DG), a local top-hat filter, or a spectral cutoff filter.}
	\label{fig:filters}
\end{figure*}

With a few exceptions, Large Eddy Simulation~(LES) methods intended for practical engineering applications are based on an implicit scale separation filter, sometimes also referred to as a grid or discretization filter. While its opposite, namely the explicit approach, allows a discretization-independent definition of the filter kernel and thus a convergence of the solution under the filter, implicit filtering typically lacks both: With the discretization inducing a local and possibly inhomogeneous cut-off length, the subfilter resolution is by design always coarse. Contrary to that, in the case of an explicit filter, its width $\Delta$ and a representative discretization spacing $h$ are chosen independently (or at least sufficiently far apart\cite{chow2003further}). For the implicit approach however, they are (approximately) identical $h\approx \Delta$. Put differently, a specific discretization induces its specific scale separation filter (both width and shape) and thus affects the resolved scales, and discretization errors interact with the represented scales. This interaction and interdependence make the closure problem for implicitly filtered LES a lot more intricate than its--already challenging--explicit counterpart. For both approaches it is a known fact that the choice of the filter kernel defines the solution and subfilter fields. \figref{fig:filters} elucidates this dependence of the LES solution and the resulting subgrid scale~(SGS) terms on the chosen filter kernel. Here, known, linear and isotropic filter kernels are applied to the same field of a Direct Numerical Simulation~(DNS), revealing the effects of that choice on the filtered field and the associated Reynolds force terms.
As also discussed by other authors~\cite{geurts2005numerically}, \citeauthor{pruett2001toward} summarizes this by concluding ``\emph{that the exact SGS-stress tensor is completely determined by the grid filter [...]}''~\cite{pruett2001toward}.
This becomes clear in \figref{fig:filters}, where the subfilter fields differ significantly for the three investigated kernels.
Thus, the closure problem for implicitly filtered LES includes both physical and numerical aspects.
In the following, we briefly discuss this dependence for simplified model equations.

The LES formulation of the governing equations and with it the definition of the closure problem is derived by coarse-graining the full-scale equations through the application of a band-width restricting filter $F(\psi)$, typically of convolution type, to a field quantity $\psi=\psi(x)$.
This yields its coarse-grained version as
\begin{align}
	\overline{\psi}(x):=F(\psi(x))=\int_\infty^\infty k(\xi-x,\Delta)\psi(\xi) d\xi,
  \label{eq:filter}
\end{align}
where $k$ is a kernel normalized such that $F(\psi_0) = \psi_0$ for a spatially constant $\psi_0$ and $\Delta$ as the filter width.
If the filter function can be expressed in the form of \eqref{eq:filter}, the filter commutes with the time derivative.
This allows to shift the filtering operation from the equation level to the solution, yielding a useful evolution equation for the coarse-grained state variables.
For the linear transport equation of a conserved quantity $u(x,t)$ with constant advection speed $a$ this yields
\begin{align}
	R(u)=\frac{\partial u }{\partial t}+a\frac{\partial u}{\partial x}=0,\nonumber\\
	\overline{R(u)}=\frac{\partial \overline{u} }{\partial t}+a\frac{\partial \overline{u}}{\partial x}=a\underbrace{\left(\frac{\partial \overline{u}}{\partial x}-\overline{\frac{\partial u}{\partial x}}\right)}_{C_1[F;\frac{\partial}{\partial x}](u)},
\end{align}
where the term $C_1$ denotes the commutation error between its arguments. As pointed out by many researchers, the commutation error $C_1$ does not vanish if $\Delta$ is nonuniform, that is if $\Delta=\Delta(x)$ or if $\Delta$ is anisotropic~\cite{geurts2006commutator,BERSELLI20071027,fureby1997mathematical,ghosal1995basic,lund2003use,john2003large,yalla2021effects}. For explicitly filtered LES, this issue occurs at domain boundaries, while for implicitly filtered LES this commutation breaks down wherever the grid spacing and thus the induced filter width changes. It is important to stress that this $C_1$ error can already occur for linear equations.

For the inviscid Burgers equation as the simplest nonlinear model, the filter operation $F$ yields the filtered equations as
\begin{align}\label{eq:burgers}
	R(u)&=\frac{\partial u }{\partial t}+\frac{1}{2}\frac{\partial u^2}{\partial x}=0,\nonumber\\
	\overline{R(u)}&=\frac{\partial \overline{u} }{\partial t}+\frac{1}{2}\frac{\partial \overline{u^2}}{\partial x}=\frac{1}{2}\underbrace{\left(\frac{\partial \overline{u^2}}{\partial x}-\overline{\frac{\partial {u^2}}{\partial x}}\right)}_{C_1[F;\frac{\partial}{\partial x}](u^2)}\nonumber\\
	&=\frac{\partial \overline{u} }{\partial t}+\frac{1}{2}\frac{\partial \overline{u}\, \overline{u}}{\partial x}=C_1+\frac{1}{2}\underbrace{\left(\frac{\partial \overline{u}\, \overline{u}}{\partial x}-\frac{\partial \overline{{u^2}}}{\partial x}\right)}_{C_2[F;()^2](u)},
\end{align}
where the classical unclosed term $C_2$ is introduced due to the quadratic flux of the nonlinear equation.

For an explicitly filtered approach, where the filter width $\Delta$ and the discretization parameter $h$ are independent and thus $h\rightarrow 0$ under the filter is possible, the formulation in \eqref{eq:burgers} is sufficient to define the closure problem.
For an implicit filter however, a discrete convective operator \smash{$\frac{\delta \overline{u}\, \overline{u}}{\delta_x}$} is introduced in addition, which induces the associated error $C_3$\cite{beck2019deep} as
\begin{align}\label{eq:burgers2}
	\overline{R(u)}=\frac{\partial \overline{u} }{\partial t}+\frac{1}{2}\frac{\delta \overline{u}\, \overline{u}}{\delta x}=C_1+C_2+\frac{1}{2}\underbrace{\left(\frac{\delta \overline{u}\, \overline{u}}{\delta x}-\frac{\partial \overline{{u}}\, \overline{{u}}}{\partial x}\right)}_{C_3[\partial_x;\delta_x]},
\end{align}
or following the notation from~\cite{geurts2005numerically} more closely as
\begin{align}\label{eq:burgers3}
	\overline{R(u)}=\frac{\partial \overline{u} }{\partial t}+\frac{1}{2}\frac{\delta \overline{u}\, \overline{u}}{\delta x}=C_1+C_2+\frac{1}{2}\frac{\partial}{\partial x}\underbrace{\left(\widehat{\overline{u}\,\overline{u}}-\overline{u}\,\overline{u}\right)}_{C_3^*[\widehat{\;.\;};\cdot]}.
\end{align}
Here, it is assumed that the discretized flux can be re-expressed as the derivative of the flux under the numerically induced filter $\widehat{\;\;\cdot\;\;}$, i.e., that \smash{$\frac{\delta}{\delta x} f= \frac{\partial}{\partial x} \hat{f}$}. %
Note that \eqref{eq:burgers2} and~\originaleqref{eq:burgers3} essentially express the same fact from a different perspective: The discretization operator induces an additional term to be closed, either directly or indirectly through its actions as a filter\footnote{A similar argument can be arrived at when considering the integral form of the equations and a finite volume discretization. Here, the divergence operator is rewritten as an integral surface flux using Gauss's divergence theorem. While the numerical integration operator and thus projection onto the mean remain exact even in a discrete setting, the integrand can become a nonlinear approximation. A typical example would be (W)ENO-type schemes. Thus, the \smash{$C_3^{(*)}$} errors also appear -- not because the integration operator in finite volume is not a box-filter with known kernel, but because its arguments have been approximated.}.

For consistent discretization schemes and smooth, bandwidth-limited fields, both $C_3$ and $C_3^*$ converge with the design order as $h\rightarrow 0$.
But these conditions are not met for the case of grid-filtered turbulence. Thus, the \smash{$C_3^{(*)}$} errors are present and can be of considerable size~\cite{geurts2005numerically}. As one contribution, they can contain a term arising from aliasing through the representation of the quadratic flux. More importantly, however, many modern, high-order accurate discretization schemes for the convective fluxes are nonlinear by design in underresolved settings to provide stability, i.e., the discretization operator itself becomes a nonlinear function of its inputs. Examples include upwinding schemes, non-oscillatory reconstructions and limiters as well as characteristics-based schemes. Hence, in practical simulations, the discrete operator $\delta_x$ (or the induced filter form $\widehat{\;\;\cdot\;\;}$) becomes a nonlinear function of the current flow solution that is not known a priori.
This has the effect that the additional terms \smash{$C_3^{(*)}$} appearing in the LES formulation are strongly dependent on the actual discretization used and cannot be determined a priori.
However, the computational subgrid stresses $C_3$ or forces $C_3^*$ can be considerably larger than the Reynolds stresses $C_2$ for $\Delta\approx h$.
Consequently, the closure problem contains both physical and numerical aspects.
However, due to the additional complexities, a lot less analysis or insight is available for the numerically induced terms. \citeauthor{piomelli1988}~\cite{piomelli1988} expressed the notion that the filter and model need to be chosen jointly, and an independent choice of both leads to decreased a priori performance. Recent developments concerning the analysis of the commutation errors $C_1$ and $C_3$ are given by \citeauthor{yalla2021effects}~\cite{yalla2021effects}, who demonstrate how to model the commutation error for differentiable filters on stretched grids and improve the SGS model accordingly. In an earlier review by \citeauthor{meneveau2000scale}~\cite{meneveau2000scale}, the authors re-express the notion that the presence of $C_1$ and $C_3$ is recognized, but no general understanding has been established. \citeauthor{ghosal1995basic}~\cite{ghosal1995basic} showed that the $C_1$ error due to a change in grid spacing is comparable to the $C_3$ error introduced by a second-order finite difference approximation and can thus be corrected by a higher-order extension. Their analysis is restricted to lossless invertible, smooth filters only -- a discretization-based filter must necessarily induce an information loss. Another interesting aspect of their analysis for the current discussion is that the $C_1$ error was found to be purely (anti-)dissipative, whereas the discretization error for the chosen case is purely dispersive. \citeauthor{vasilyev2004local}~\cite{vasilyev2004local} also found that the properties of the commutation error can vary considerably with the filter type, which supports the notion that the commutation errors are a function of the specific discretization scheme and that $C_1$ and $C_3$ must be considered jointly. Moreover, \citeauthor{geurts2006commutator}~\cite{geurts2006commutator} showed that the overall commutation error has the same order of magnitude and scaling with the filter width as the turbulent stress fluxes themselves. They make the argument that thus not just $C_2$ should be modeled, but the other commutators should be parameterized as well. Still, due to the increased complexity of considering additional unclosed terms and their dependence on the discretization parameters, even recent works state that ``\emph{[...] the commutation error arising from the implicit part of the filter (i.e., projection onto the underlying discretization) has not been well investigated.}"~\cite{yalla2021effects}.
Thus, $C_1$ and \smash{$C_3^{(*)}$} are often left unclosed due to lack of knowledge about them and the major focus in LES model development is to find an approximation for $C_2$.

In conclusion, in implicitly filtered LES, discretizations induce an intrinsic scale separation filter that induces a filtered solution and unclosed fields that are both specific to the employed discretization.
This affects the contributions to the closure problem and adds an error that is a function of the specifics of the discretization. Expressed differently, the filter kernel enacted by the discretization is typically neither uniform nor linear, which makes its analysis beyond simplified cases intractable and probably even more nonuniversal than for the Reynolds stresses themselves. This fits to the often observed fact that (presumably) identical closure models give diverging results when combined with different discretizations~\cite{beck2016influence}.

In this work, we address this issue and present a novel approach for finding optimal closure strategies for implicitly filtered LES, i.e., a discretization-consistent (or filter-consistent) closure model. It is based on the formulation of an in-the-loop optimization problem that adapts or extends existing closure models. From the discussion above it has become clear that an a priori definition of the closure terms and thus of an appropriate model is difficult to impossible, as the discretization-induced filter is nonstationary, inhomogeneous and possibly nonlinear.
Therefore, the optimization of such closure approaches must occur in conjunction with the LES operator itself in a dynamic fashion, since a segregated approach, where an optimization is carried out over independently generated prior data, is not constructive\cite{beck2019deep,kurz2021investigating}. Instead, we solve this task through Reinforcement Learning~(RL) on the LES level and demonstrate how to find optimal explicit and implicit closure models that are specific to the discretization and match or outperform the current state of the art. We build on our previous work~\cite{kurz2022relexi,kurz2023deep} but extend the discussion by two completely novel aspects: The focus on discretization-consistent SGS models and an implicit modeling strategy based on a hybrid discretization operator presented in \secref{sec:implicit_closure}. We proceed from here as follows.

In \secref{sec:sota}, we review the current literature on discretization adapted or optimized closure models, followed by a short introduction to the optimization via reinforcement learning in \secref{sec:rl}. In \secref{sec:les}, the baseline discretization schemes and the LES formulations are presented by introducing the explicit and implicit closure models in \secref{sec:explicit_closure} and \secref{sec:implicit_closure}, respectively
\secref{sec:setup} provides details on the LES test case and the employed optimization scheme. In \secref{sec:training}, we present an overview of the training process and and of the optimization results achieved. A detailed analysis for the explicit and implicit modeling approaches follows in \secref{sec:results_explicit} and\secref{sec:results_implicit}, respectively.
Finally, the major findings of the work are summarized and some potential next steps are outlined in \secref{sec:conclusion}.

\section{Optimization-based Closure Models}
\label{sec:optclosure}

\subsection{State of the Art}
\label{sec:sota}

As discussed above, the theoretical derivation of closure models for implicitly filtered LES is typically challenging due to the strong interaction between the filtered nonlinear equations and the employed nonlinear discretization.
To this end, it is common practice to derive turbulence models in the context of the explicitly filtered equations, where the effects of the applied discretization on the filtered description vanish or are sufficiently separated in wavenumber space.
These models are then also applied in the implicit case. 
To account for the differences between the contexts they were originally derived for and the one they are now deployed in, the empirical parameters of such models typically have to be calibrated in order to yield satisfactory results, see for example the thorough study by~\citeauthor{meyers2006optimal}~\cite{meyers2006optimal}.
This fitting of model parameters yields the most simplistic case of optimization in terms of turbulence modeling.
The most prevalent example of this is the tuning ``by hand'' of the parameter $C_s$ of the Smagorinsky model~\cite{matheou2016numerical,VIRE20111903,meyers2006optimal}. 
In the context of implicit subgrid modeling, it seems natural to understand the choice and optimization of a model as a function of the discretization.
Especially the idea of using nonlinearly stable discretization schemes for the convective fluxes and their nonlinear dissipation as main element for the modeling has shown considerable success~\cite{adams2007relation,flad2017use}.

An example of how the discretization operators can be adapted systematically in this approach was proposed by~\citeauthor{hickel2006adaptive}~\cite{hickel2006adaptive}, who used an evolutionary algorithm to optimize the free parameters of a WENO-type discretization scheme in an automated fashion to optimize the schemes' truncation error as an implicit subgrid model. %
Similarly, \citeauthor{schranner2016optimization}~\cite{schranner2016optimization} employed a WENO-type scheme, where the free parameters are optimized with a response surface approach such that the numerical truncation error mimics the closure terms.
\citeauthor{flad2020large}~\cite{flad2020large} presented a similar approach for DG methods by applying the Nelder–Mead method to optimize the coefficients of a modal relaxation filter for a Discontinuous Galerkin method to recover the turbulent statistics of a Homogeneous Isotropic Turbulence~(HIT) flow.

More recently, \citeauthor{maulik2020spatiotemporally}~\cite{maulik2020spatiotemporally} employed an optimization with ML based on Supervised Learning~(SL) to train an Artificial Neural Network~(ANN) as classifier that decides whether distinct points in the computational domain require numerical dissipation.
Based on this either a central dissipation-free discretization or a upwinding scheme is employed, where the introduced dissipation accounts for the required closure.
As an alternative learning paradigm, RL trains an agent by interactions with a dynamical environment; thus, this intrinsic connection to dynamical systems has first led to a strong interest in the use of this ML method for flow control~\cite{rabault2019artificial,rabault2019accelerating,tang2020robust,fan2020reinforcement,varela2022deep,vignon2023recent,vinuesa2022flow}.
However, the same properties make RL a highly interesting optimization framework for the task of finding discretization-consistent closure models or optimized discretization operators.
First attempts for RL-based closures were reported by \citeauthor{novati2021automating}~\cite{novati2021automating} and \citeauthor{kurz2023deep}~\cite{kurz2023deep}, who employed RL to derive optimal explicit closure by optimizing the model coefficient of a standard Smagorinsky model dynamically in space and time.
Similarly, \citeauthor{kim2022deep}~\cite{kim2022deep} used RL to approximate the subgrid stresses in channel flows and first applications of RL for wall-modeling are reported by Bae and colleagues~\cite{bae2022scientific,vadrot2023log}. The idea of optimizing the discretization itself towards specific dispersion and dissipation properties through RL has recently been shown to be a promising route\cite{FENG2023112436}.
Based on these initial successes, the RL paradigm appears to be a promising candidate for optimization-based closures in LES, as it is suited for highly nonlinear, dynamical systems and can embed complex, nonlocal decision strategies.
In this work, RL is thus the optimization method of choice to find discretization-consistent closure models.
To this end, the following section provides a brief introduction to RL.

\subsection{Reinforcement Learning}
\label{sec:rl}

\begin{figure}
	\centering
  \ifthenelse{\boolean{final}}{
    \includegraphics[width=0.99\linewidth]{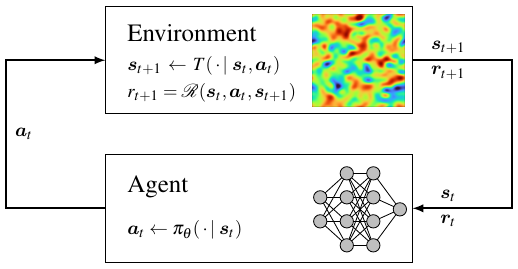}
  }{
    \tikzsetnextfilename{fig_tikz_mdp}
    \input{tikz/mdp.tikz}
  }
  \caption{General outline of a Markov Decision Process (MDP) as the main problem formulation in RL.}
  \label{fig:rl}
\end{figure}

RL is a field of machine learning that aims at finding the optimal behavior of an agent interacting with some dynamical environment.
The basic outline of an RL task is shown in \figref{fig:rl}.
Based on the current state of the environment~$\ppvec{s}_t$, the agent can perform an action $\ppvec{a}_t$ that causes the environment to transition from the current state~$\ppvec{s}_t$ to a new state~$\ppvec{s}_{t+1}$.
This state transition follows its transition probability function~$\ppvec{s}_{t+1} \leftarrow T(\,\cdot\,|\, \ppvec{a}_t,\ppvec{s}_t)$, which gives the conditional probability of transitioning to a distinct new state~$\ppvec{s}_{t+1}$ given the current state~$\ppvec{s}_{t}$ and action~$\ppvec{a}_{t}$.
Based on the tuple~$(\ppvec{s}_t,\ppvec{a}_t,\ppvec{s}_{t+1})$ the transition is assigned a scalar reward~$r_{t+1}$ that is determined by the reward function as
\begin{equation}
  r_{t+1} = \mathcal{R}(\ppvec{s}_t,\ppvec{a}_t,\ppvec{s}_{t+1}).
\end{equation}
The reward function acts as a performance metric such that a higher reward corresponds to a more beneficial state transition or, to put it differently, a more beneficial action taken by the agent.
The agent then performs a new action~$\ppvec{a}_{t+1}$ based on the new state~$\ppvec{s}_{t+1}$, until some terminal state~$\ppvec{s}_n$ is reached.
This process is called a Markov Decision Process~(MDP).

The actions an agent performs are determined by its policy~$\pi_{\theta}=\pi_{\theta}(\ppvec{a}_t |\,\ppvec{s}_t)$, which describes the probability of performing action~$\ppvec{a}_t$ given the current state of the environment~$\ppvec{s}_t$.
In the context of deep RL, this policy is represented by a deep ANN with~$\theta$ as the ANN's weights.
The agent's goal in such an RL problem is then to collect the maximum amount of reward for the given MDP, which corresponds to finding the optimal policy~$\pi_{\theta}^*$ or rather its optimal set of parameters~$\theta^{*}$ that maximize the expectation.
Over the last decade, a plethora of different RL methods has been proposed that all introduce some algorithm to iteratively optimize the policy's weights based on sampled interactions between the agent and the environment to find a suitable approximation of the optimal policy~$\pi_{\theta}^*$.
In this work, the Proximal Policy Optimization 
(PPO)~\cite{schulman2017proximal} algorithm is employed to train the agent, which comprises two phases that are performed iteratively.
First, a batch of interactions are sampled from the environment based on the current policy and second, the policy is optimized with the sampled experience using the policy gradient theorem~\cite{sutton2018reinforcement}.
These phases are repeated until the model converges.
For more details on RL and PPO, the reader is referred to the standard work by \citeauthor{sutton2018reinforcement}~\cite{sutton2018reinforcement} and the excellent summary by \citeauthor{notter2021hierarchical}~\cite{notter2021hierarchical}, respectively.

\section{LES Formulations and Closures}
\label{sec:les}

In this section, first, the baseline discretization scheme for LES is introduced in~\secref{sec:baseline_les_scheme}.
The chosen discretization is based on a DG scheme and acts as the implicit scale-separating filter for all LES carried out in this work.
Based on this, two closure approaches are presented.
First, a well-known explicit closure approach is detailed in \secref{sec:explicit_closure} and second, a novel implicit closure scheme is introduced in \secref{sec:implicit_closure}.
Both closure approaches are conceptually very different and are chosen deliberately to investigate the flexibility and generalizability of the optimization approach for both model classes.

\subsection{Baseline LES scheme}
\label{sec:baseline_les_scheme}

\begin{figure}
  \ifthenelse{\boolean{final}}{
    \includegraphics[width=0.55\linewidth]{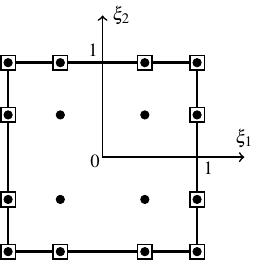}
  }{
    \tikzsetnextfilename{fig_tikz_gausspoints_lgl}
    \input{tikz/gausspoints_lgl.tikz}
  }
  \caption{Illustration of the DGSEM reference element in two dimensions. The dots represent the LGL interpolation nodes for $N=3$ and the squares indicate where the surface fluxes between the elements and its neighbors are evaluated through numerical flux functions.}
  \label{fig:dg_solution}
\end{figure}

All simulations in this work are performed using the Discontinuous Galerkin Spectral Element Method~(DGSEM)~\cite{kopriva2009implementing} implemented in the flow solver FLEXI (available under \url{https://github.com/flexi-framework/flexi}). This family of methods has seen a widespread interest and use in the community for DNS and LES of especially compressible turbulence~\cite{FERRER2023108700,WITHERDEN20143028,krais2021flexi}.
For the DGSEM, the domain is decomposed into a set of arbitrarily structured, nonoverlapping elements.
Each element is then mapped into the reference element $E\in[-1,1]^3$ illustrated in \figref{fig:dg_solution} using a polynomial mapping $\ppvec{x}= \ppvec{\chi}(\ppvec{\xi})$, where $\ppvec{\xi}=(\xi_1,\xi_2,\xi_3)$ and $\ppvec{x}=(x_1,x_2,x_3)$ denote the three-dimensional coordinates in the reference and physical space, respectively.
The solution is represented by a polynomial of degree $N$ expressed in a basis which is defined by the tensor product of one-dimensional Lagrange polynomials, which yields
\begin{equation}
  \ppvec{U}(\ppvec{\xi},t) \approx \sum_{i,j,k=0}^N \hat{\ppvec{U}}_{ijk}(t)\ell_i(\xi_1)\ell_j(\xi_2)\ell_k(\xi_3),
\end{equation}
where $\ell_i(\xi_{1,2,3})$ denotes the $i$-th one-dimensional Lagrange polynomial defined with respect to a specific set of interpolation nodes and $\hat{\ppvec{U}}_{ijk}(t)$ are the time-dependent solution coefficients.
The DGSEM is based on the weak form of the governing equations, which can be derived by projection of the residual onto the test space on the reference element $E$.
The spatial projection operators are discretized by a Gauss-type quadrature associated with the chosen set of interpolation and integration points.
Throughout this work, Legendre--Gauss--Lobatto (LGL) nodes are used for interpolation and integration. Their distribution in the reference element is illustrated exemplary in \figref{fig:dg_solution} for $N=3$.
The individual elements are coupled via the numerical flux over the element faces.
However, since two neighboring elements are not required to share the same solution at the faces but can rather exhibit different solutions at the shared edge as illustrated in \figref{fig:fv_blending}, the resulting flux at the shared edge is not unique.
To this end, (approximate) Riemann solvers are employed to compute a single unique numerical flux for both elements.
This highlights the hybrid character of a DG discretization, which resembles a finite-element-type discretization with local projection within each element, where the solution is represented by a high-order, continuous ansatz function, and Riemann solvers as known from finite-volume (FV) methods are used to account for the discontinuities at the element faces.
The reader is referred to the standard work by \citeauthor{kopriva2009implementing}~\cite{kopriva2009implementing} for more details.

The semi-discrete form results in an ODE in time that can be written as
\begin{equation}
  \frac{\partial}{\partial t}
  \hat{\ppvec{U}}
  =
  \ppvec{R}^{DG}\left(\hat{\ppvec{U}}(t)\right)
  ,
  \label{eq:strong_form}
\end{equation}
and is integrated in time using a suitable integration scheme.
To counter aliasing errors that arise in underresolved simulations, a split-form DG formulation\cite{gassner2016split,pirozzoli2011numerical} is employed to ensure the nonlinear stability of the scheme.
The unique Euler fluxes at the element boundaries are computed using the L2-Roe Riemann solver\cite{osswald2016l2roe} and the gradients for the parabolic fluxes of the Navier--Stokes equations are computed using the first lifting method of \citeauthor{bassi1997high}~\cite{bassi1997high} often referred to as the ``BR1'' method.
Extensive details on the methods, its properties, the implementation and validation for LES and other flows are given in~\cite{gassner2013accuracy,hindenlang2012explicit,krais2021flexi}.

\subsection{Explicit closure scheme}
\label{sec:explicit_closure}
The easiest and arguably most widespread class of explicit closure schemes relies on Boussinesq's hypothesis, which assumes that the closure terms are proportional to the local shear stress.
This allows to model the effect of the closure terms as an eddy viscosity that can be added directly to the physical viscosity.
While this hypothesis is usually questionable for the general case~\cite{meneveau2000scale,klemmer2023turbulence}, such eddy viscosity models enjoy immense popularity due to their simplicity to implement, their intuitive interpretability and their great success in practical applications.
One of the first, and arguably still one of the most prominent, eddy viscosity models is the standard Smagorinsky model (SSM), which computes the eddy viscosity at a given point in space as
\begin{equation}
  \mu_{SGS}
  =
  \overline{\rho}\left(C_s\,\Delta\right)^2
  \sqrt{2 S_{ij} S_{ij}}
  ,\quad\text{with}\;\;
  i,j=1,2,3,
  \label{eq:smago}
\end{equation}
with $C_s$ as an empirical model coefficient and $\ppvec{S}$ as the rate-of-strain tensor computed from filtered quantities, which is defined as
\begin{equation}
  S_{ij}
  =
  \frac{1}{2}\left( \frac{\partial u_i}{\partial x_j} + \frac{\partial u_j}{\partial x_i}\right).
  \label{eq:rateofstrain}
\end{equation}
As before, $\Delta$ denotes the filter width of the low-pass LES filter.
This filter width is not uniquely defined in a multi-dimensional and unstructured case, but is typically computed in the DG context as
\begin{equation}
  \Delta = \frac{\sqrt[3]{Vol}}{N+1},
\end{equation}
where $Vol$ denotes the volume of a DG element.
The filter width is additionally normalized by the $N+1$ degrees of freedom in each direction within the DG element to make it comparable to finite-difference and finite-volume methods.
It is important to stress that this choice of a filter length is generally not unique, since the solution points and also the respective filter width are not equally distributed across the element, but rather differ throughout the DG element.
Moreover, the suitability of the computed isotropic filter width from the element's volume becomes increasingly questionable the more elongated and skewed grid elements become.

To this end, a straightforward extension of the standard model is to adapt the model coefficient in space to account for these differences in filter width.
This can be done on two levels.
First, the model coefficient $C_s$ can be adapted individually for each DG element, as typically done for the dynamic Smagorinsky model.
However, second, one can also account for the inhomogeneities within each DG element by allowing the model coefficient to vary within each element, i.e., $C_s = C_s(\ppvec{\xi})$.

In the context of DG, this approach can be implemented by representing $C_s(\ppvec{\xi})$ as a polynomial with free coefficients in each element, represented in a modal Legendre basis, which yields
\begin{equation}
  C_s(\ppvec{\xi}) = \sum_{i,j,k=0}^N \mathcal{C}_{s,ijk} \phi_i(\xi_1) \phi_j(\xi_2) \phi_k(\xi_3),
  \label{eq:legendre_basis}
\end{equation}
where $\mathcal{C}_{s,ijk}$ is the tensor comprising the coefficients and \smash{$\phi_i(\xi_{1,2,3})$} denotes the $i$-th one-dimensional Legendre polynomial in the $\xi_1$-,$\xi_2$- and $\xi_3$-direction, respectively.
In a first step, we only consider the constant and quadratic mode, which yields
\begin{equation}
  \mathcal{C}_{s,ijk}
  =
  \begin{cases}
    C_s^{const} & \text{if}\; i=j=k=0\\
    C_s^{quad}  & \text{if}\; i=2 \wedge j=k=0\\
    C_s^{quad}  & \text{if}\; j=2 \wedge i=k=0\\
    C_s^{quad}  & \text{if}\; k=2 \wedge i=j=0\\
    0 & \text{otherwise}\\
  \end{cases}
  .
\end{equation}
Hence, only four nonzero values are set.
The constant contribution for the zeroth mode is set to $ C_s^{const}$ and the quadratic coefficients in each of the three directions are set to the same value $C_s^{quad}$, i.e., we consider a shifted hyperparaboloid of revolution.
This choice is justified for isotropic closure problems like the driven HIT introduced in Sec.~\ref{sec:testcase}.
Note that choosing $C_s^{quad}=0$ results in the classical formulation with a constant $C_s$; here however per grid element.
To compute the eddy viscosity at each solution point following \eqref{eq:smago}, the nodal value of $C_s$ at each solution point can be computed with
\begin{equation}
  C_{s,mno},
  =
  V_{mi}V_{nj}V_{ok}
  \mathcal{C}_{s,ijk},
  \quad
  \text{with}
  \quad
  V_{ij} = \phi_i(\xi^{(j)}),
\end{equation}
where $\ppvec{V}$ denotes the one-dimensional interpolation matrix that interpolates the Legendre basis to the Lagrange basis used to represent the solution.
The entries of $\ppvec{V}$ can be obtained by evaluating the Legendre basis functions~$\phi_i(\xi)$ at the LGL nodes~\smash{$\{\xi^{(j)}\}_{j=0}^N$}.
Note that this case recovers the elementwise constant case for~\smash{$C_s^{quad}=0$}.
The idea of a varying eddy viscosity is of course well-known, for instance from spectral vanishing viscosity models.
Here, instead of formulating the eddy viscosity as a function in wavenumber space, we define it as a function of physical space.
Thus, the explicit model in this work is based on a classical Smagorinsky eddy viscosity formulation, where the model coefficient~$C_s$ can be chosen either as a constant in each grid element or as an individual shifted parabolic polynomial generated by the zeroth and second order Legendre modes.
This leads to the two modeling scenarios and optimizable parameters in each grid element:
\begin{equation}	\label{eq:cschoice}
  \begin{aligned}
    \text{RL-Const:} &\quad C_s^{const}\in \mathds{R} ,\, C_s^{quad}=0 ,\\
    \text{RL-Quad:}  &\quad C_s^{const}\in \mathds{R} ,\, C_s^{quad}\in \mathds{R}.
  \end{aligned}
\end{equation}

\subsection{Implicit closure scheme}
\label{sec:implicit_closure}

\begin{figure}[t]
  \ifthenelse{\boolean{final}}{
    \includegraphics[width=0.99\linewidth]{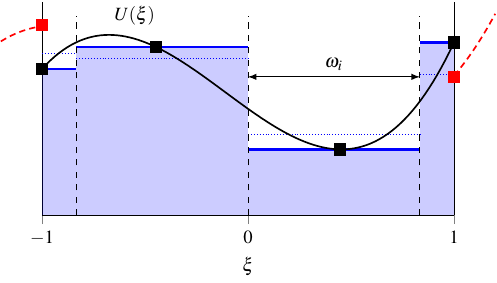}
  }{
    \tikzsetnextfilename{fig_tikz_dgfv}
    \input{tikz/dgfv/dgfv.tikz}
  }
  \caption{Outline of the reference element for the hybrid DG/FV blending scheme in one dimension. The DG solution polynomial $U(\xi)$ is given in black with the LGL interpolation points highlighted as squares. The integral mean values of the compatible FV scheme are highlighted in blue with the dotted line indicating the theoretically exact integral mean for a given subcell. The size of the subcells corresponds to the LGL integration weights $\omega_i$. The solutions of the left and right neighbor elements are indicated in red.}
  \label{fig:fv_blending}
\end{figure}

\begin{figure*}[t]
	\centering
  \includegraphics[width=0.48\linewidth]{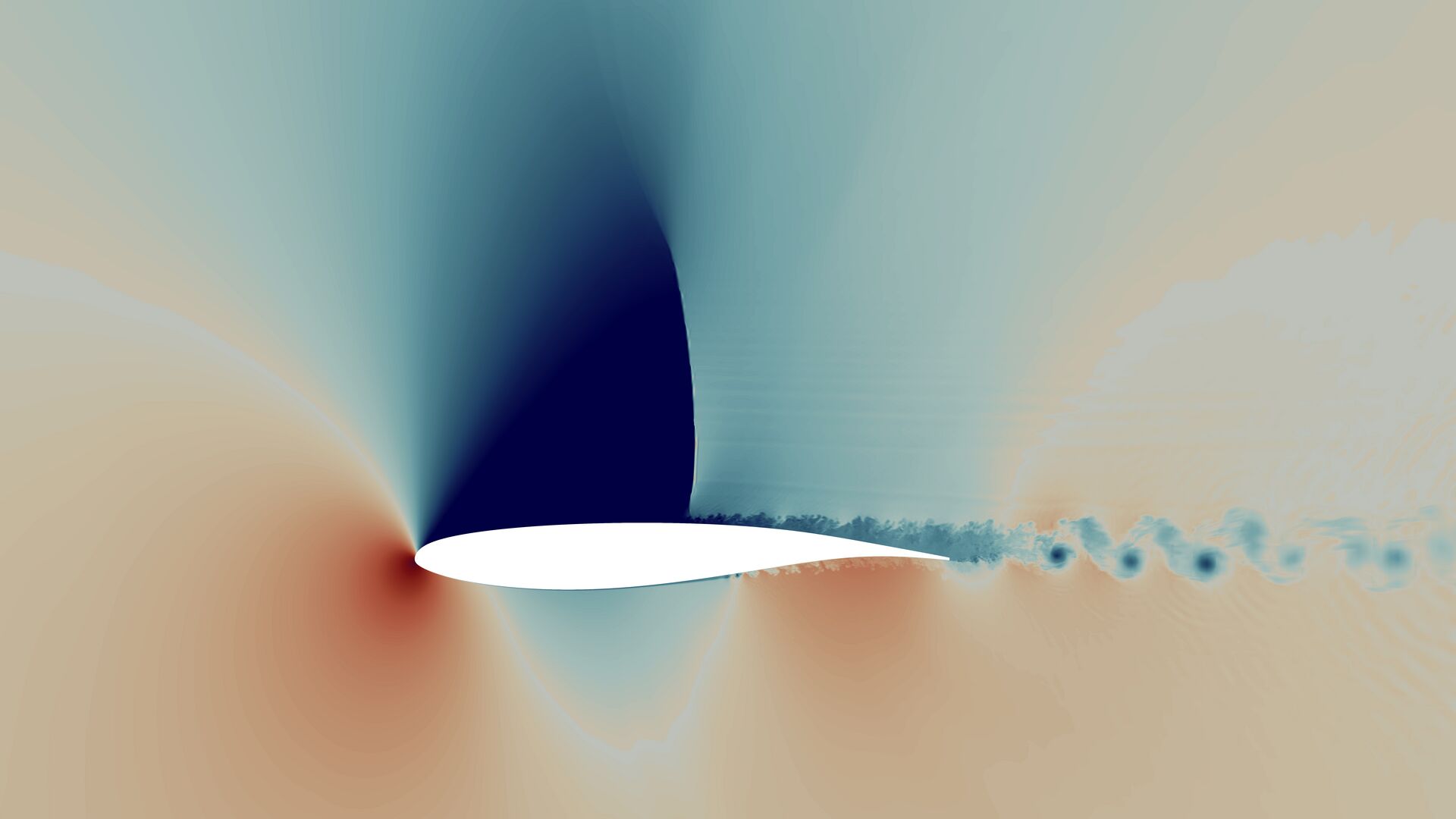}
  \hfill
  \includegraphics[width=0.48\linewidth]{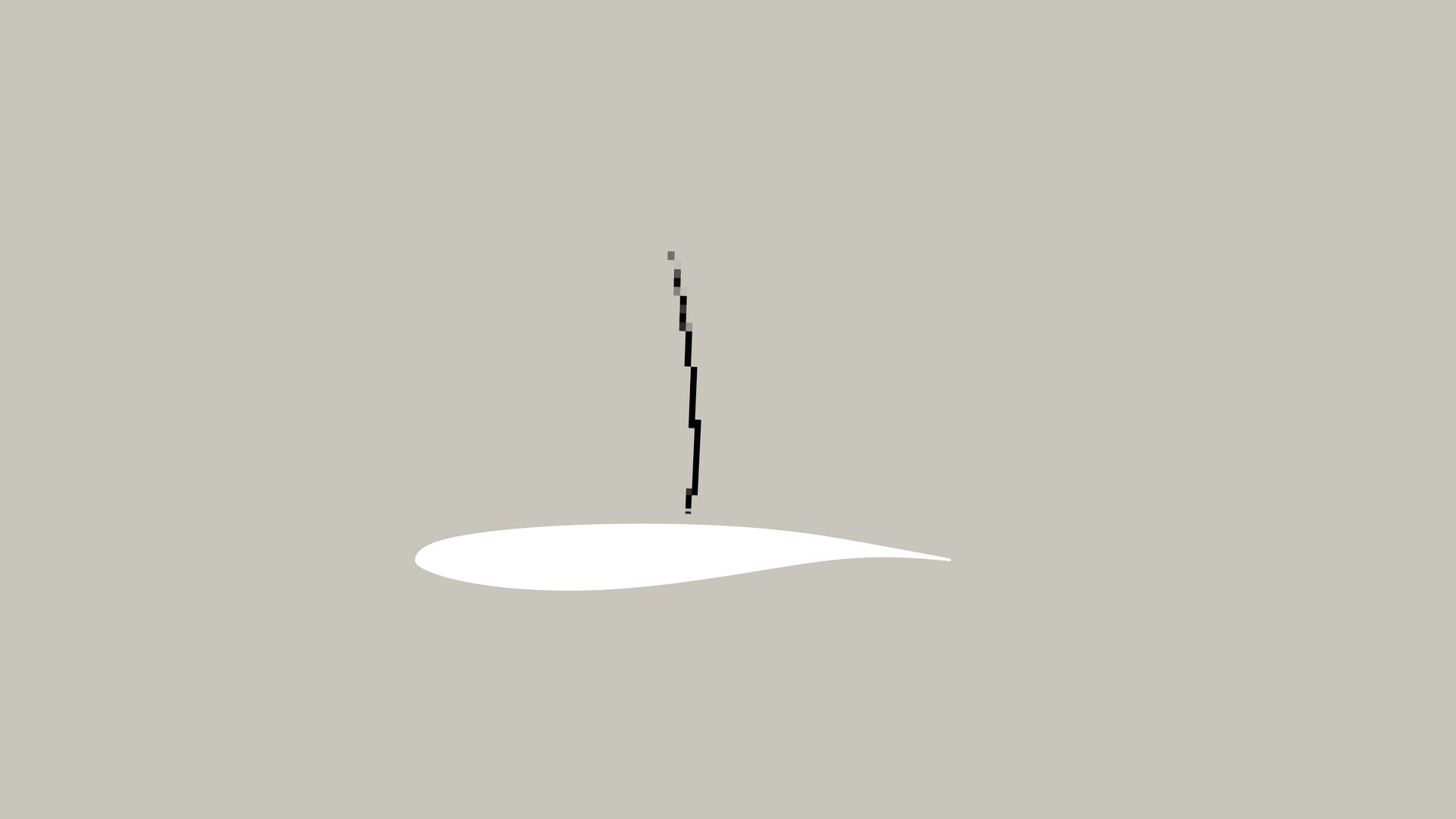}
  \caption{Transonic flow around an OAT15A profile with an emerging shock on the suction (upper) side. Left: Two-dimensional slice of the instantaneous field solution colored by the density. Right: Scaling parameter $\alpha$ with $\alpha=1$ appears as black region near the shock and $\alpha=0$ in the rest of the domain.}
	\label{fig:OAT_fv_blending}
\end{figure*}

In the implicit modeling case, the closure terms are not approximated explicitly. Instead, it is acknowledged that the nonlinear solution scheme and its numerical dissipation act as an implicit dissipation mechanism.
\citeauthor{boris1992new}~\cite{boris1992new} recognized that second-order, monotone FV schemes can act as such a source of damping and thereby induce a low-pass filter on the solution.
Since then, implicit LES has been adopted widely for all sorts of schemes, including DG methods~\cite{flad2017use,uranga2011implicit}.
As discussed more detailed in \secref{sec:sota}, implicit turbulence modeling can be summarized as optimizing the truncation error of the numerical scheme to yield a suitable approximation of the true closure terms.
In the DG context, this manipulation of the truncation terms has in the past been interpreted as choosing a certain polynomial degree, adaptive polynomial de-aliasing, setting a suitable Riemann solver or using a for instance structure preserving formulations for the convective fluxes.

In this work, we propose a different and novel approach and strive to optimize the discretization scheme more directly for implicit LES using RL.
For this, the hybrid DG/FV blending approach proposed by \citeauthor{hennemann2021provably}~\cite{hennemann2021provably} is applied.
This hybrid scheme was originally developed as a shock capturing algorithm for high-order DG methods, since they tend to oscillate near strong gradients such as shocks, which can cause simulations to become unstable.
To this end, a compatible low-order FV scheme that possesses the Total-Variation-Diminishing (TVD) property is applied in such regions to stabilize the overall scheme.
An example of this original field of application is shown in \figref{fig:OAT_fv_blending}, where the transonic flow over a OAT15 airfoil causes a shock to emerge on the suction side.
This shock is detected and addressed using the hybrid DG/FV scheme to yield a stable and accurate simulation with the goal to preserve the high-order accuracy of the method in the turbulent flow regions, while exploiting the TVD property and dissipation of FV to stabilize the solution near the shock. 

The blending is performed for each DG element as a convex combination between the high-order DG operator $\ppvec{R}^{DG}(\hat{\ppvec{U}})$ and the compatible low-order FV subcell method $\ppvec{R}^{FV}(\hat{\ppvec{U}})$, which yields the semi-discrete form as
\begin{equation}
  \frac{\partial}{\partial t} \hat{\ppvec{U}} = (1-\alpha)\ppvec{R}^{DG}(\hat{\ppvec{U}}) + \alpha \ppvec{R}^{FV}(\hat{\ppvec{U}}),
  \label{eq:fvblending}
\end{equation}
with $\alpha\in[0,1]$ as the blending parameter.
Clearly, the DG method can be recovered from \eqref{eq:fvblending} as a special case for $\alpha=0$, while a pure FV method follows for $\alpha=1$.
Since both discretizations for themselves are conservative and consistent, the convex combination of both operators can also be shown to be conservative and consistent~\cite{hennemann2021provably}.
The hybrid DG/FV scheme requires some form of indicator to determine which regions require FV stabilization as shown in \figref{fig:OAT_fv_blending}.
\citeauthor{hennemann2021provably} themselves proposed an \emph{a priori}~\cite{hennemann2021provably} indicator to determine a suitable value for $\alpha$ in each element during the simulation, which was later complemented by colleagues with an \emph{a posteriori} indicator~\cite{rueda2021subcell,rueda2022subcell}.
However, other indicators are also possible and especially data-driven indicators based on ML present themselves as promising candidates, as seen for instance in the works of \citeauthor{beck2020neural} \cite{beck2020neural}, \citeauthor{schwarz2023reinforcement}~\cite{schwarz2023reinforcement} and \citeauthor{ray2018artificial}~\cite{ray2018artificial,ray2019detecting}.
As an extension, we strive to employ an RL-based optimization to find a suitable control strategy to adapt $\alpha$ for implicit LES in this work.

This gives the following model scenario with a free parameter in each grid element: 
\begin{equation}	\label{eq:alphachoice}
	\begin{aligned}
    \text{RL-Blend:  }&\quad \left\{\alpha \in \mathds{R} \;|\; 0 \le \alpha \le 1 \right\}.
	\end{aligned}
\end{equation}
It is important to note that the blending is local to each specific element and that all elements can be treated independently from each other.
This follows from the coupling fluxes across the element faces, which are identical for the DG and FV method, since in both cases the same numerical fluxes are computed based on the same solution at the element's face.
While originally proposed for improved shock capturing, the concept of blending a high-order accurate discretization with little, localized dissipation and a TVD scheme draws on the same idea for SGS modeling through such operators proposed by~\citeauthor{boris1992new}~\cite{boris1992new}.
Thus, such blended schemes seem worth investigating in the context of implicit LES modeling, especially for compressible flows. 

\section{Test Case and Training Setup}
\label{sec:setup}

\subsection{Test Case}
\label{sec:testcase}

\begin{figure*}
  \begin{minipage}[C]{0.45\textwidth}
    \includegraphics[width=0.99\textwidth]{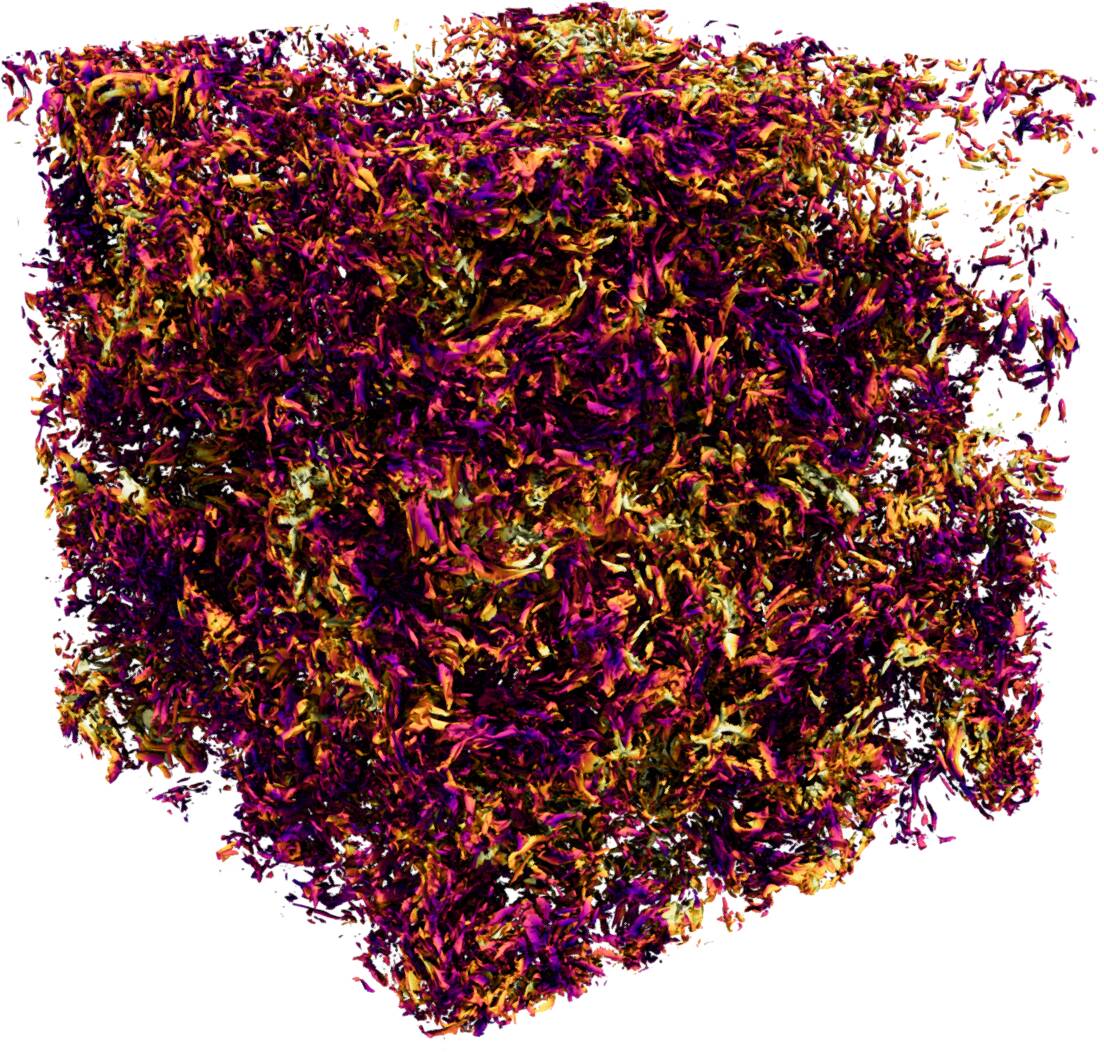}
  \end{minipage}
  \hfill
  \begin{minipage}[C]{0.45\textwidth}
    \ifthenelse{\boolean{final}}{
      \includegraphics[width=0.99\linewidth]{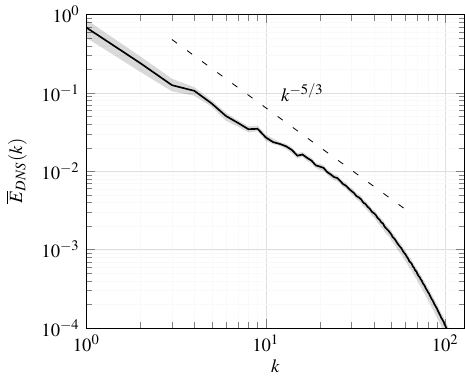}
    }{
      \tikzsetnextfilename{fig_tikz_spectra_DNS}
      \input{tikz/fhit/fig_spectra_DNS.tikz}
    }
  \end{minipage}
  \caption{\emph{Left}:~Instantaneous flow field of the forced HIT visualized by iso-surfaces of the $\lambda_2$-criterion colored by the velocity magnitude. \emph{Right}:~Mean TKE spectrum with the shaded area indicating the minimum and maximum observed instantaneous values.}
  \label{fig:fhit}
\end{figure*}

Since this work is focused on exploring the feasibility of the novel RL-optimized approach, we restrict the investigation to the most basic test bed for turbulent flows.
For this, we use a forced Homogeneous Isotropic Turbulence~(HIT) to evaluate the performance of the different modeling strategies.
This test case can be considered as \emph{turbulence-in-a-box} and allows an unobstructed evaluation of the LES approaches for the building block of all turbulence.
The computational domain is a cube of size $\ppvec{x}\in[0,2\pi]^3$ with periodic boundary conditions, in which a random turbulent flow field is initialized following Rogallo's procedure\cite{rogallo1981numerical}.
These initial fluctuations break down over time into real turbulent flow as illustrated in \figref{fig:fhit}.
The physical viscosity in the system causes the initial amount of Turbulent Kinetic Energy~(TKE) to dissipate over time.
To this end, an isotropic forcing algorithm~\cite{lundgren2003linearly,de2015anisotropic} is employed to inject the amount of dissipated energy back into the system to obtain a quasi-steady turbulent flow state with stationary turbulent statistics.
It is important to note that the applied forcing acts on a nodal level and thus injects energy not only at the lowest wavenumbers, as would be the case in general turbulence, but instead at all wavenumbers present in the solution.
This causes the slight shift in slope in comparison to the theoretical slope of $k^{-5/3}$ that can be observed in \figref{fig:fhit} and was already recognized in the literature~\cite{lundgren2003linearly,de2015anisotropic}.
Based on this setup, a DNS of a forced HIT at a Reynolds number of $\mathrm{Re}_{\lambda}\approx 180$ with respect to the Taylor microscale is computed with the flow solver FLEXI as a reference and ground-truth for the different investigated LES modeling approaches.
For the DNS a polynomial degree of $N=7$ is employed, which results in an eighth-order scheme.
The computational mesh comprises $64^3$ elements which results in a total resolution of \num{512} interpolation points in each spatial direction and thus in a total of 134 million degrees of freedom~(DOF) per solution variable.
The DNS is computed up to a nondimensional time of $t=10$ corresponding to around 14 large eddy turnover.
The simulation was run on \num{16384}~CPU cores on HAWK at the High-Performance Computing Center Stuttgart~(HLRS) and took almost 2 hours of wall time resulting in a total cost of around \num{30000}~CPU-hours.
Based on this DNS setup, the LES used to evaluate the performance of the different modeling strategies employ either 24~DOF and 36~DOF per spatial direction and a polynomial degree $N$ of either 3, 5 or 8. Note that this also entails a change in the induced filter shape of the discretization, as a $N=3$ scheme has for example a broader and stronger spectral dissipative footprint.
Generally, the higher the order of the ansatz, the larger the ratio of resolved to represented scales becomes.
For the schemes used in this work, an analysis of the dispersion and dissipation properties gives an estimate for the cut-off wavenumber of 4 points per wavelength~(ppw)~\cite{gassner2011comparison}.
Thus, spectral content discretized by less than 4~ppw can no longer be considered resolved, but only represented.
Note that the usual cut-off number~$k_{max}$ given by the Nyquist frequency of 2~ppw is the theoretical limit for Fourier-based discretizations but is unachievable with a general polynomial ansatz.
With these considerations in mind, we note that the chosen resolution of 24~DOF and 36~DOF are low for the LES at this Reynolds number, with the cut-off wavenumber barely reaching into the inertial range.
This makes this scenario particularly challenging for the closure strategies and brings out the details between the different models and discretizations more sharply.
All LES simulations were initialized from appropriately filtered DNS fields.
All reported results and comparison runs are computed based on the same initial flow field.

\subsection{Optimization Setup and Training}
\label{sec:training_setup}
Put into the context of an LES and the task of finding an optimal closure model, the MDP from \secref{sec:rl} can be specified as follows.
An individual, implicitly-filtered LES with the chosen numerical scheme and closure strategy provide the environment.
The state $\ppvec{s}_t$ of the environment is the flow solution at time~$t$ and the action~$\ppvec{a}_{t}$ chosen by the ANN embedding the policy~$\pi_{\theta}$ is one of the choices of \eqref{eq:cschoice} and \eqref{eq:alphachoice}.
At each time step and in each grid element, a closure model according to the action $\ppvec{a}_{t}$ is solved alongside the usual integration process, advancing the solution for a time interval~$\Delta t_{RL}$ to obtain the next flow solution~$\ppvec{s}_{t+1}$.
This cycle repeats until the end time of the LES is reached.
The metric that determines the success of the optimization (and of the solution of the MDP through the policy) is the expected cumulative reward during the training iterations.
The transferability of the found, optimal policy to a new, unseen flow case must then be established in a second evaluation step.

In this work, the reward is defined via the mean squared error of the TKE spectrum of the LES denoted by~$E_{LES}(k)$ and the temporally averaged DNS spectrum~$\overline{E}_{DNS}(k)$.
In addition, an exponential is used to normalize the reward to the interval~$r_t=[-1,1]$, which yields the reward function as
\begin{equation}
  r = 2\, \exp\left(\frac{\beta}{k_{max}}\sum_{k=1}^{k_{max}}\left(\frac{\overline{E}_{DNS}(k)-E_{LES}(k)}{\overline{E}_{DNS}(k)}\right)^2\right)-1.
  \label{eq:reward}
\end{equation}
Here, $k_{max}$ is the maximum wavenumber that is considered and $\beta$ is an empirical scaling parameter.
Throughout this work, the 24~DOF cases are trained with $k_{max}=9$ and the 36~DOF cases with $k_{max}=11$, while all configurations use a scaling factor of $\beta=0.4$.

We deem it important to clarify that, while the DNS results are available in this case, the RL paradigm does also allow to incorporate a broader range of prior knowledge into the optimization target such as theoretical findings, data from experiments or implicit constraints such as numerical stability.
In the context of turbulence modeling, this prior definition of a performance metric is not new, but rather also the common approach for analytical models.
Here, the performance of different models is assessed and compared by applying them to canonical test cases, where a high-fidelity solution is known, and comparing how well certain high-level statistics are recovered by the models.
The training process of the RL problem is thus comparable to the process of tuning the empirical parameters of an analytical closure model to optimize its performance on such canonical reference cases, which is common practice in turbulence modeling~\cite{pope2009turbulent}.

This problem setup is implemented within the Relexi framework~(\url{https://github.com/flexi-framework/relexi}), which allows to run simulation codes as training environments on high-performance computing hardware as detailed in~\cite{kurz2022relexi,kurz2023deep,kurz2022deep}.
The flow simulations used as training environments are computed using FLEXI as flow solver and the PPO~\cite{schulman2017proximal} is employed as training algorithm to optimize the agent based on the obtained interactions with the environment.
The employed policy ANN consists of 5 layers of three-dimensional convolutions with ReLU activation functions and a total of around \num{3300} parameters as detailed by \citeauthor{kurz2023deep}~\cite{kurz2023deep}.
However, in contrast to this previous work, the actions predicted follow a beta distribution instead of a Gaussian, which has shown to accelerate the training~\cite{chou2017improving}.
To this end, the policy yields as output the two parameters of a beta distribution for each output quantity, from which the respective action can be sampled.
For this, the output of the last layer $x$ is followed by an activation function that reads as
\begin{equation}
  y = \underbrace{\log(1+\exp(x))}_{\mathrm{softplus}(x)}+1
\end{equation}
to ensure that all parameters of the beta distribution are retained within the range $(1,\infty)$.
Since the beta distribution itself is defined on the interval $[0,1]$, the drawn values are transformed from this standard interval to the ranges shown in \tabref{tab:variable_range} using a linear transformation.
For the RL-Quad model, the final $C_s$ values resulting from \smash{$C_s^{const}$} and \smash{$C_s^{quad}$} are restricted to positive values using $C_s \leftarrow \max\{C_s,0\}$.

\begin{table}
  \begin{tabular}{lcccccc}
    \toprule
    Variable && $C_s^{const}$ && $C_s^{quad}$   && $\alpha$  \\
    \midrule
    Interval && $[0,0.5]$     && $[-0.25,0.25]$ && $[0,0.5]$ \\
    \bottomrule
  \end{tabular}
  \caption{Intervals the different predicted variables are transformed to from the standard interval $[0,1]$ of the beta distribution.}
  \label{tab:variable_range}
\end{table}

In all cases, the initial states for the training runs are obtained by filtering flow states of the DNS to the respective LES resolution.
Based on these initial states, the simulations are advanced for 5 nondimensional time units, which corresponds to around 7 large-eddy turnover times.
The policy updates the current actions every $\Delta t_{RL}=0.1$ time units, which results in 50 sampled interactions per episode.
The computational cost per LES was at most 0.8 CPUh for the 36~DOF blending simulations, while all other configurations were considerably cheaper.
For the 24~DOF case, each LES was computed on 8 CPU cores each, while 16 CPU cores were used for each LES for the 36~DOF configurations.
In each training iteration 16 episodes are sampled in parallel and 5 epochs of optimization on the full training data are performed for the actor and critic networks using the Adam optimizer~\cite{kingma2014adam} with a learning rate of $10^{-4}$.
The training is performed for \num{2000} iterations for each configuration.

For details on the coupling of our LES solver and the RL agent, the details of the RL method, the parallelization and previous results we refer the reader to our previous work~\cite{kurz2022deep,kurz2022relexi,kurz2023deep}.

\section{Results}
\label{sec:results}
First, the general success of the training for the different configurations detailed in \secref{sec:training} is discussed in \secref{sec:training}.
Based on this, the results for the explicit modeling approach are reported in \secref{sec:results_explicit} and for the implicit closure scheme in \secref{sec:results_implicit}.
As reference, the results are always compared to the reference DNS described in \secref{sec:testcase} as well as three commonly used LES closure strategies.
This comprises first, the classic Smagorinsky model with a global, static $C_s=0.17$~(SSM)\cite{smagorinsky1963general} and second, its dynamic variant~(DSM) according to \citeauthor{moin1991dynamic}~\cite{moin1991dynamic} with implementation details given in previous work~\cite{kurz2023deep}.
  It is important to note that the theoretical value of $C_s=0.17$ that is chosen here is in principle only valid for specific explicit LES filters.
 Moreover, the value considered optimal varies depending on the specific filter employed, the type of flow and, in the case of implicitly filtered LES, the applied discretization~\cite{pope2009turbulent}.
  Since the optimal value for a given case is thus generally not known a priori, the choice $C_s=0.17$ acts as a plausible and general point of reference that is applied for all considered configurations.
  A more detailed discussion of the SSM in the context of DG can be found in \citeauthor{beck2016influence}~\cite{beck2016influence}.
We also include the no-model LES formulation~(NoMo), where the discretization errors of the baseline scheme are acting as the only closure term.
This approach is often used in the context of high-order DG discretization schemes and has shown its validity for a number of cases~\cite{beck2014high,flad2017use,FERRER2017754,BASSI2016367,MOURA2017615,dewiart}.
It is often included in the category of implicit models, as no additional, explicitly computed closure terms are included.
In order to distinguish it from ``true'' implicit modeling, we label it no-model LES here.

All flow quantities presented in the following are obtained by initializing the LES with the unseen evaluation state and computing the LES up to $t=20$.
Only the flow solution within $t\in[10,20]$ is then considered and averaged to obtain the results.
This ensures that the LES are long-term stable and become sufficiently independent of the initial state.

In addition to the results for Cartesian meshes presented in the following, results for the inference of the trained models on deformed meshes are given in Appendix~\ref{app:deformed}.

\subsection{Overview}
\label{sec:training}

\begin{figure*}
  \ifthenelse{\boolean{final}}{
    \includegraphics[width=0.99\linewidth]{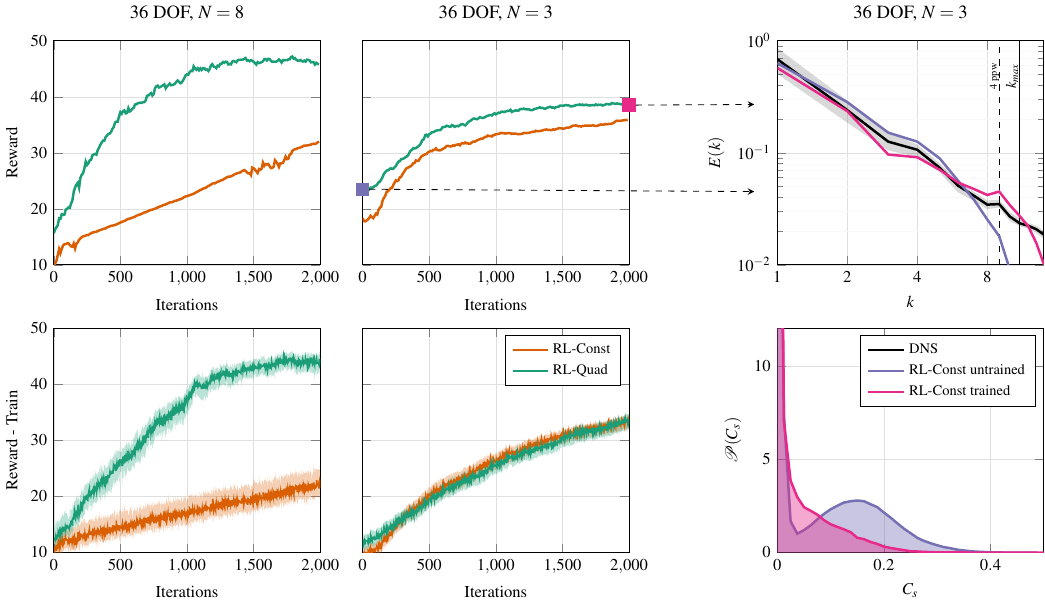}
  }{
    \tikzsetnextfilename{fig_tikz_training_reward}
    \input{tikz/fig_training_reward.tikz}
  }
  \caption{Training results for the explicit closure models. \emph{Left and center:} Evolution of the collected reward during the training on the evaluation~(\emph{top}) and training runs~(\emph{bottom}), where the shaded area indicates the minimum and maximum return and the line the average. \emph{Right:}~Results of the evaluation run of the randomly initialized and the trained policy for RL-Const and $N=3$ showing the TKE spectra~(\emph{top}) and the corresponding histogram of predicted $C_s$ values~(\emph{bottom}). The squares in the respective colors emphasize in the reward plot on the left which policy corresponds to the spectra and histograms on the right.}
  \label{fig:results_overview}
\end{figure*}

We start by giving a general overview of the optimization process.
\figref{fig:results_overview} \emph{left} exemplary details the evolution of the collected reward for the explicit closure models (\eqref{eq:cschoice}) over the training iterations for the configurations using 36 DOF in each spatial direction and both $N=3$ and $N=8$.
For both cases, the top row shows the collected reward when starting on an unseen initial state of the HIT (evaluation reward), while the bottom row corresponds to the reward collected during training (training reward).
At the beginning of the training, the applied model is initialized with random weights and thus achieves only little reward.
During the training, the collected reward along a simulation episode then increases steadily with each training iteration until the collected reward converges to a maximum.
It is striking that the reward collected during the evaluation runs is consistently larger than the training reward. This can be due to two reasons.
First, the evaluation run is initialized with an unseen initial state that is not used for the training.
Coincidently, this state could be a more favorable initialization for the policy than the ones used during training.
Secondly and more importantly, the policy is evaluated greedily in the evaluation run, i.e., the action with the highest expected reward is always chosen.
During training, each action is drawn at random from the prescribed beta distribution to facilitate exploration.
This discrepancy illustrates that the beta distribution allows to employ quite different train and evaluation policies in order to allow for sufficient exploration, while yielding accurate results if evaluated greedily during the evaluation runs. %

For both cases in~\figref{fig:results_overview} \emph{left}, i.e., $N=3$ and $N=8$, the model with quadratically distributed $C_s$ parameter (RL-Quad) outperforms the elementwise constant model (RL-Const) in terms of the collected reward.
This is to be expected since the RL-Const is a special case of the RL-Quad and thus acts as a lower bound in terms of performance. Our results confirm this.
It is however interesting that the difference between both models is considerably smaller for the $N=3$ case than for the $N=8$ case, where the RL-Const only improves quite slowly and does not reach a converged state by the end of the training.
We presume that the reason behind this is that the quadratic distribution of eddy viscosity can adapt better to the inhomogeneities of the underlying discretization operator and thus closure terms.
This notion is supported by the fact that the difference in reward between RL-Const and RL-Quad increases with $N$, which we also confirmed for $N=5$.
When keeping the total amount of DOF constant, an increase of $N$ implies a reduced number of grid elements per direction and thus a larger potential difference between a constant and a quadratic distribution in each element. It also increases the inhomogeneity of the discretization operator, thus also increasing the potential improvement for a policy giving a local $C_s = C_s(\ppvec{\xi})$.
We note that the training in \figref{fig:results_overview} is not converged in all cases, thus, this is still an open question requiring confirmation.
Comparing the performance of $N=8$ and $N=3$ overall, the higher ansatz cases achieve also consistently higher rewards.
This can be attributed to the increased scale-resolving capabilities and thus lower numerical error of the higher-order methods, resulting in a better approximation of $\delta_x \approx \partial_x$ and thus a reduced error term $C_3$ in \eqref{eq:burgers2}.

\begin{figure*}
  \ifthenelse{\boolean{final}}{
    \includegraphics[width=0.95\linewidth]{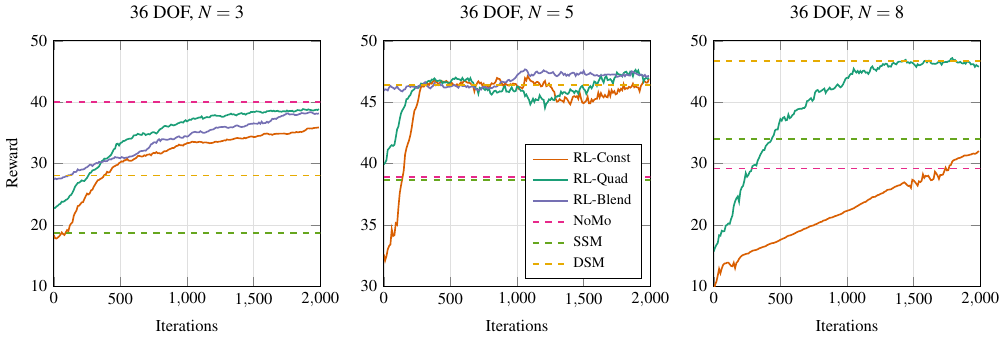}
  }{
    \tikzsetnextfilename{fig_tikz_training_reward_comp}
    \input{tikz/fig_training_reward_comp.tikz}
  }
  \caption{Evolution of the reward collected in the evaluation runs during training for different configurations. The RL-Const and RL-Quad results refer to the explicit closure cases discussed in \secref{sec:results_explicit}, the RL-Blend case to \secref{sec:results_implicit}. The rewards that would be collected by the NoMo, SSM and DSM cases are provided as a reference.}
	\label{fig:reward_comp}
\end{figure*}

The right column in \figref{fig:results_overview} illustrates the effect of the training.
The top figure shows the spectra of TKE for the 36~DOF, $N=3$ case alongside the reference DNS.
The untrained RL case corresponds to iteration 0 in the reward plots and the trained case to the converged policy at iteration 2000.
In accordance with the increase in reward, the quality of the LES spectra improves for the trained RL, which gives significantly better results up to the cut-off wavenumber.
The bottom figure shows the PDF of the $C_s$ values predicted by the untrained and trained policy during the course of an LES.

The randomly initialized policy yields a distribution that resembles a Gaussian about a mean of about 0.17 with its right tail squeezed near zero to fit into the interval $C_s\in[0,0.5]$.
Its proposed $C_s$ leads clearly to too much $\mu_{sgs}$ and result in the spectrum shown above with an energy blockage in the lower wavenumbers and the drop in the higher ones.
After training, the PDF is skewed stronger towards lower values for $C_s$, which results in overall less dissipation and a better fit of the TKE spectrum. Details will be discussed in the coming sections.\

\figref{fig:reward_comp} shows the evolution of the collected reward during the training for the configurations discussed in the following sections.
Generally, all investigated cases for explicit and implicit closures are able to improve in the course of the training and reach a (at least approximately) converged return at the end of the training.
The notable exception here is the configuration using the Smagorinsky model with an elementwise constant parameter (RL-Const) for $N=8$, which is not converged, as already discussed above.
For comparison, the rewards that would be obtained for the three reference closures (NoMo, SSM, DSM) are also given. It is notable that their performance clearly depends on the combination with the specific discretization and that there is no unverisally best model, as discussed in the introduction. This observation again supports the argument that the closure model and the discretization must be considered jointly.

We observe here that the rewards for the NoMo decrease with $N$.
This is expected since the dissipation in the baseline scheme is introduced through the coupling fluxes the elements' surface.
For higher $N$, the numerical dissipation thus becomes increasingly localized at the elements' faces.
For the $N=3$ case, the NoMo LES gives very accurate results as also reported in literature, even surpassing the DSM, which gives the highest rewards for the other cases. This reduced performance of the DSM can be attributed to the breakdown of the scale invariance assumption and thus the validity of the test filtering used in the dynamic procedure of the DSM, as for this low resolution, the cut-off wavenumber is barely touching the inertial range.
For both the SSM and the DSM, the achieved reward rises with $N$, and the overall best performance also increases accordingly.
This can be attributed to the improved resolution capability of the baseline scheme, which provides a more accurate spectral representation to the closure model and hence, also reduces the demands on it.

For our investigations it is important to note that all RL-based results achieve a very high reward that is close to or even higher than the best among the three classical models.
\figref{fig:reward_comp} clearly shows that the RL-based models are able to outperform the DSM for the $N=5$ and $N=8$ case, where the DSM yields the highest reward among the analytical models.
In contrast, the no-model LES demonstrates the best performance of the reference models for the $N=3$ case, which cannot be reached by the RL models.
While the RL-based models thus do not yield the best performance for each specific case, the RL approach yields competitive results for all cases and arguably the best overall performance for the range of cases investigated in this study.
We can also observe that the RL-Quad results consistently give slightly higher rewards than the RL-Const case.

Based on these preliminary observations, the following section provides a more detailed discussion of the obtained turbulence models and their performance on the given test case.
Since the $N=8$ results have not fully converged, we will focus on the $N=3$ and $N=5$ cases in the following.

\subsection{Explicit Closure}
\label{sec:results_explicit}

\begin{figure*}
  \ifthenelse{\boolean{final}}{
    \includegraphics[width=0.81\linewidth]{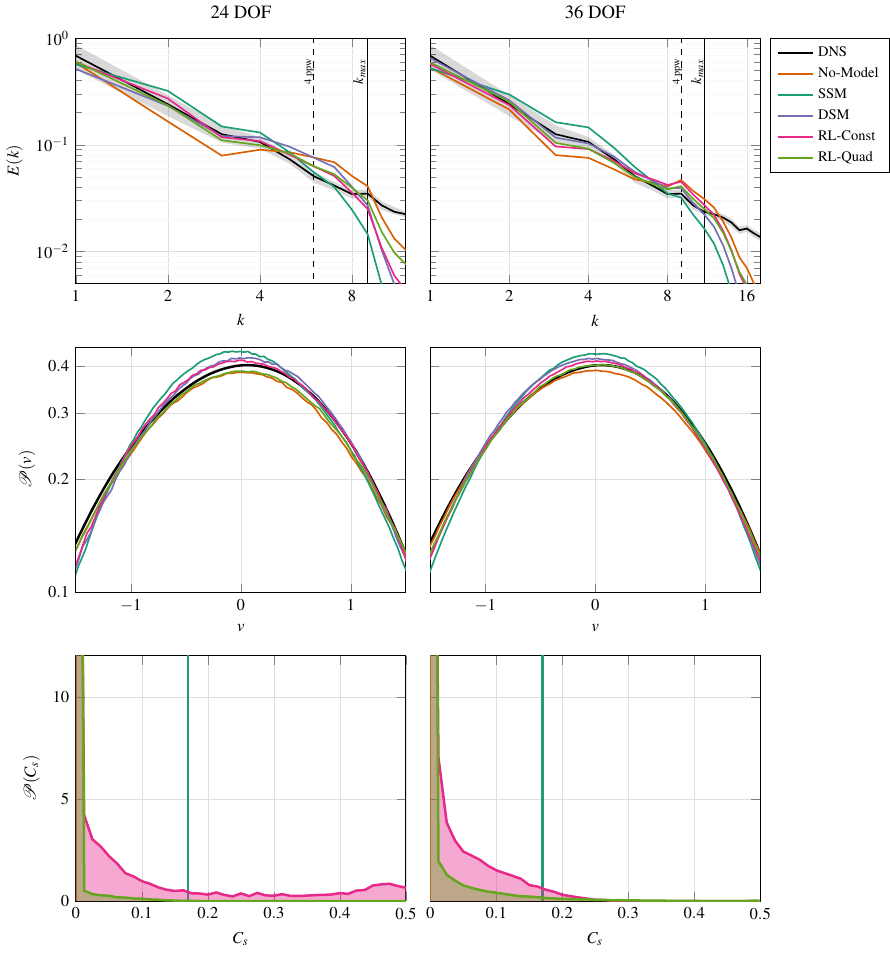}
  }{
    \tikzsetnextfilename{fig_tikz_spectra_smago}
    \input{tikz/fig_spectra_smago.tikz}
  }
  \caption{Results for the explicit RL closures predicting either a constant (RL-Const) or quadratic distribution (RL-Quad) of the $C_s$ parameter within each element for the resolution with 24~DOF (\emph{left}) and 36~DOF (\emph{right}) per spatial direction with $N=5$. The results for the NoMo, SSM, DSM and the DNS solution are shown for reference with the shaded area indicating the minimum and maximum observed TKE per wavenumber in the DNS. \emph{Top}: TKE spectrum, \emph{middle}: PDF of the velocity fluctuations, \emph{bottom}: PDF of the $C_s$ values predicted by the RL models.}
	\label{fig:cs_comp_DOFs}
\end{figure*}

This section provides a detailed discussion of the results obtained for the explicit eddy viscosity-type closure strategies introduced in \secref{sec:explicit_closure}.
\figref{fig:cs_comp_DOFs} shows the results for the same baseline scheme $N=5$ and two resolutions of 24 and 36 DOF for both explicit RL-based models, the three reference closures strategies and the reference DNS.
In the top row, the TKE spectra for the no-model LES shows the typical behavior for an underdissipative method with a buildup of energy around the cut-off and a drainage of the medium wavenumbers.
Opposite to that, the SSM induces a too severe damping of the high frequency content and thus a blockage at the medium range.
Both RL models are in excellent agreement with the reference DNS in the complete useful wavenumber range, with the results of the RL-Quad model showing a slight advantage.
Both in the reward as in the quantities shown here, the DSM results for the 36 DOF case are very close to the RL models, and all of them are again in very good agreement with the reference.
For the DSM, this very good performance for the selected test case is not surprising\cite{vollant2016dynamic}, but it is remarkable that the optimization can find a strategy for selecting a good $C_s$ dynamically on the resolved field alone without making the assumption of scale invariance or access to a test-filtered solution.
For the case with 24~DOF, the resolution becomes marginal for a proper LES.
While the RL-Const and RL-Quad solutions retain their excellent agreement to the DNS, the DSM results show more pronounced deviations.
While we have not investigated the reason for this, it is plausible that due to the marginal resolution, the test filter is no longer situated in the similarity region and the basic assumption of the DSM becomes questionable.
The distributions of the velocity fluctuations (middle row) confirm these findings that the No-Model and SSM give mediocre results, while the DSM and both RL-models are in very good agreement with the DNS.

\begin{figure*}
  \ifthenelse{\boolean{final}}{
    \includegraphics[width=0.79\linewidth]{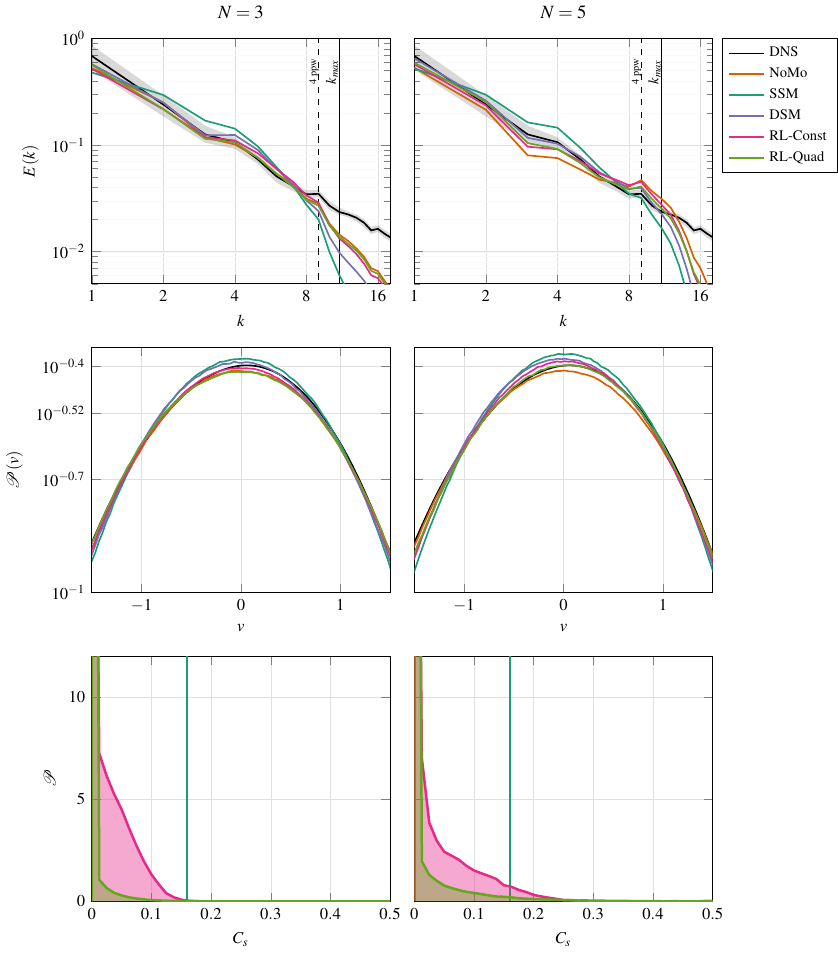}
  }{
    \tikzsetnextfilename{fig_tikz_spectra_smago_N3N5N8}
    \input{tikz/fig_spectra_smago_N3N5N8.tikz}
  }
  \caption{Results for the explicit RL closures predicting either a constant (RL-Const) or quadratic distribution (RL-Quad) of the $C_s$ parameter within each element for the resolution with 36~DOF and polynomial degrees $N=3$ or $N=5$. The right column corresponds to the right column in \figref{fig:cs_comp_DOFs} and is shown here again for better comparability. The results for the no-model LES, the SSM, the DSM and the reference DNS solution are shown for reference with the shaded area indicating the minimum and maximum observed TKE per wavenumber in the DNS. \emph{Top}: TKE spectrum, \emph{middle}: PDF of the velocity fluctuations, \emph{bottom}: PDF of the $C_s$ values predicted by the RL models.}
	\label{fig:cs_comp_N}
\end{figure*}

\begin{figure*}
  \ifthenelse{\boolean{final}}{
    \includegraphics[width=0.77\linewidth]{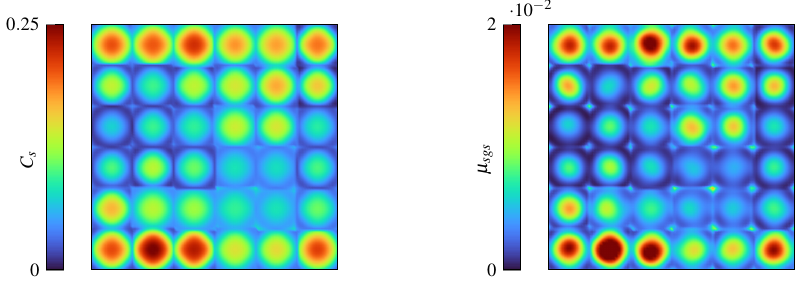}
  }{
    \tikzsetnextfilename{fig_tikz_smago_quad_field}
    \input{tikz/smago_quad_field/fig_smago_quad_field.tikz}
  }
  \caption{Two-dimensional slice of the trained agent's predictions for the explicit closure model with the quadratic distribution of $C_s$ for the 36 DOF case with $N=5$. The slice is extracted from a distinct position in the domain to illustrate the distribution in the elements' center and then averaged over the time interval $t\in[10,20]$. \emph{Left}: distribution of the $C_s$ in the elements' center. \emph{Right}: resulting distribution of eddy viscosity $\mu_{SGS}$ in the temporal mean.
  \label{fig:cs_distribution}}
\end{figure*}

The bottom row of \figref{fig:cs_comp_DOFs} shows the distribution of the actions proposed by the RL models for the LES runs.
It should be stated that ML-based models are notoriously hard to interpret, but analyzing their behavior allows to draw at least some initial conclusions.
The first observation here is that all RL models give nontrivial distributions with strong positive skew, which means that for most instances in space and time the model picks little or no viscosity.
The second observation is that the mean for the RL-Const case is higher than for the RL-Quad case, which supports the notion that for the RL-Quad case, the eddy viscosity can be targeted better to the underlying discretization (c.f. the discussion on \figref{fig:cs_distribution}).
Lastly, for the RL-Const case, the distribution is wider for the lower resolution case of 24~DOF.
This fits the notion that more dissipation is needed since less of the TKE is resolvable in that case.

Summing up the discussion of \figref{fig:cs_comp_DOFs}, the learned strategies select eddy viscosity constants that significantly improve the results over the SSM case, even for a marginal LES resolution.
We note that these results are consistent with our previous findings~\cite{kurz2023deep}, where a Gaussian distribution for the action space was used instead of a beta distribution with finite support here.

In a next step, the \figref{fig:cs_comp_N} compares the results for a fixed resolution of 36 DOF and varying $N$, which tests the capabilities of the RL-based approach to adapt to different discretization properties and induced filters.
As already discussed for \figref{fig:reward_comp}, the no-model results are remarkably good for the $N=3$ case, but deviate for higher $N$.
For both discretizations, the RL-based models are in excellent agreement with the reference data, with the RL-Quad model showing a slightly better fit.
Given that the no-model approach is quite accurate for $N=3$, the $C_s$ distribution proposed by the RL-agent with a high positive skew can be interpreted as mimicking this $C_s=0$ behavior of the no-model case.
For $N=5$, where the inherent dissipation induced by the scheme itself is insufficient, the optimizer proposes a less skewed distribution with a higher mean value for $C_s$, which results in a much better agreement with the reference in the TKE spectra.
These results indicate that the RL-based approach is capable of adjusting the optimal closure strategy to different discretization schemes and produces consistently accurate, discretization-consistent closure models over varying resolutions and $N$ when trained accordingly.

This feature is investigated further for the RL-Quad model in the $N=5$, 36 DOF case, for which the time-averaged proposed $C_s$ and the resulting eddy viscosity $\mu_{SGS}$ are shown on a 2D slice through the domain in \figref{fig:cs_distribution}.
Here, the individual $6\times 6$ grid elements are clearly visible.
Since the policy is evaluated individually for each element, the resulting predictions within each element is independent from the others.
Remarkably, on average, the proposed strategy of the optimization yields a parabolic distribution of $C_s$ with the same orientation across all elements.
This causes the maximum to reside in the element center and the eddy viscosity to reduce towards zero at the boundaries.
As described in \secref{sec:explicit_closure}, we specifically added the quadratic degree of freedom to the inner-element $C_s$ distribution to account for inhomogeneities in the DG-operator itself, i.e., in $C^*_3$.
These inhomogeneities are a consequence of the hybrid FV/FE character of the DG scheme, where numerical dissipation is introduced only through the numerical fluxes at the element boundaries.
The optimal eddy viscosity found here shows an inverting behavior: No additional $\mu_{SGS}$ at the boundaries but added volume-dissipation within the element.
This is a strong indication that the RL-agent has found a closure strategy that not just accounts for the closure terms $C_2$ as usual but has also taken the numerical properties of the scheme into account while doing so.
We note that this indicates that this approach of considering both the numerical and model dissipation in conjunction can lead to a more homogeneous overall operator, but this requires additional investigations.

\subsection{Implicit Closure}
\label{sec:results_implicit}

\begin{figure}
  \ifthenelse{\boolean{final}}{
    \includegraphics[width=0.84\linewidth]{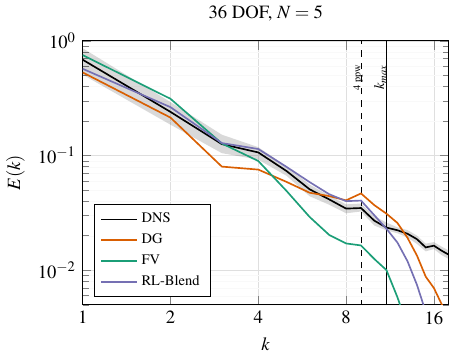}
  }{
    \tikzsetnextfilename{fig_tikz_spectra_blend_alpha_comp}
	  \input{tikz/fig_spectra_blend_alpha_comp.tikz}
  }
	\caption{Comparison of the TKE spectra of no-model LES computations for the 36~DOF, $N=5$ case using either a pure DG scheme~($\alpha=0$), a pure FV scheme~($\alpha=1$) or the RL-informed blending approach.
  \label{fig:alpha0_1}}
\end{figure}

In a next step, we investigate the implicit closure consisting of choosing a convex blending of the DG and FV discretization operators as introduced in \secref{sec:implicit_closure}.
It is important to stress that this hybrid scheme was originally proposed for shock capturing and has not been interpreted as a possible implicit SGS model alternative yet.
However, the notion of a nonlinear viscosity operator builds the basis for a number of implicit modeling approaches as summarized in \secref{sec:sota}.
\figref{fig:alpha0_1} shows the no-model LES for the baseline pure DG scheme ($\alpha=0$), the pure FV scheme ($\alpha=1$) and a blended variant with RL-optimized $\alpha$.
The pure DG solution is underdissipative, while the pure FV solution is overly dissipative.
However, the RL optimization scheme identifies a suitable blending strategy that results in a hybrid DG/FV scheme that matches the DNS spectrum very accurately.
A plausible explanation for this is that, similar to the RL-Quad case discussed in \secref{sec:results_implicit}, the dissipation induced by the FV operator is inverse to the dissipation of the DG scheme.
This is due to the distribution of FV subcells shown in \figref{fig:fv_blending}, where the larger subcells in the element's center introduce more dissipation than the smaller subcells near the faces.
This is in stark contrast to the dissipation properties of the DG method, where the numerical dissipation is introduced solely by the numerical fluxes at the faces.
Blending both operators could thus again result in a more homogeneous distribution of dissipation.

\begin{figure*}[t!]
  \ifthenelse{\boolean{final}}{
    \includegraphics[width=0.85\linewidth]{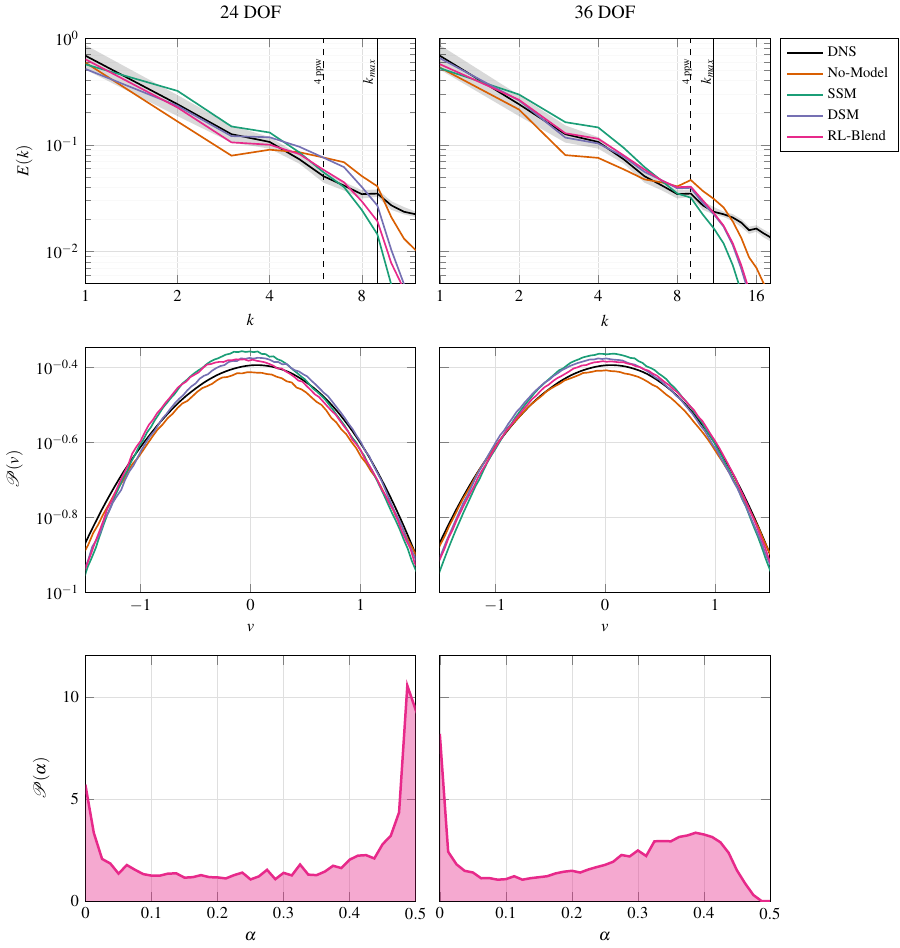}
  }{
    \tikzsetnextfilename{fig_tikz_spectra_blend}
    \input{tikz/fig_spectra_blend.tikz}
  }
  \caption{Results for the implicit RL closure (RL-Blend), which predicts a blending factor $\alpha$ within each element for the resolution with 24~DOF (\emph{left}) and 36~DOF (\emph{right}) per spatial direction with $N=5$. The results for the no-model LES, the SSM, the DSM and the reference DNS solution are shown for reference with the shaded area indicating the minimum and maximum observed TKE per wavenumber in the DNS. \emph{Top}: TKE spectrum, \emph{middle}: PDF of the velocity fluctuations, \emph{bottom}: PDF of the $\alpha$ values predicted by the RL models.}
  \label{fig:blending_DOF}
\end{figure*}

The accumulated training rewards in \figref{fig:reward_comp} indicate that the blending approach can indeed gather considerable reward that is comparable to that of the explicit RL-based closures.
This is confirmed in \figref{fig:blending_DOF}, which demonstrates the excellent agreement of the TKE spectra of the RL-Blend approach particularly for the higher resolution case.
The distributions of the blending factor in the bottom row of \figref{fig:reward_comp} exhibit again a nontrivial behavior, where a higher $\alpha$ corresponds to the stronger contribution of the (dissipative) FV scheme, as follows from \eqref{eq:fvblending}.
Consistent with the reduced resolution in the 24~DOF case, the blending factor tends to a higher FV-contribution.

\begin{figure*}[t!]
  \ifthenelse{\boolean{final}}{
    \includegraphics[width=0.85\linewidth]{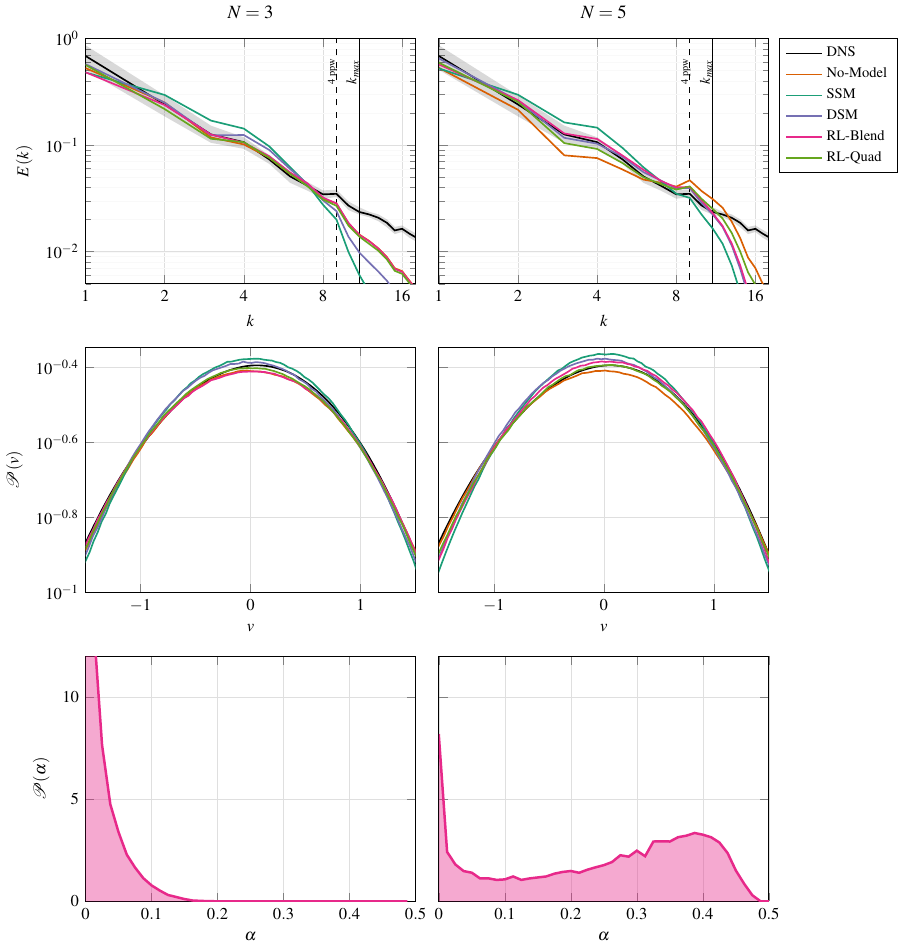}
  }{
    \tikzsetnextfilename{fig_tikz_spectra_blend_N3N5N8}
    \input{tikz/fig_spectra_blend_N3N5N8.tikz}
  }
  \caption{Results for the implicit RL closure (RL-Blend), predicting a blending factor $\alpha$ within each element for the resolution with 36~DOF and polynomial degrees $N=3$ and $N=5$. The right column corresponds to the right column in \figref{fig:blending_DOF}, with the results for RL-Quad added for comparability. The results for the no-model LES, the SSM, the DSM and the reference DNS solution are shown for reference with the shaded area indicating the minimum and maximum observed TKE per wavenumber in the DNS. \emph{Top}: TKE spectrum, \emph{middle}: PDF of the velocity fluctuations, \emph{bottom}: PDF of the $C_s$ values predicted by the RL models.}
  \label{fig:blending_N}
\end{figure*}

\figref{fig:blending_N} shows the results for different polynomial degrees $N$ and compares them to the RL-Quad model, which performed best from the two explicit RL models.
It is remarkable that both of these RL-based models, RL-Quad and RL-Blend, show very similar and highly accurate results for both $N$ investigated, although both are based on very different modeling ideas.
For the $N=3$ case, where the no-model solution (corresponding to a pure DG scheme with $\alpha=0$) already gives very good results, the trained agent mimics this behavior and reduces the amount of FV blended into the scheme, which results in a distribution that is skewed highly to the left.
In contrast, for the $N=5$ case the numerical dissipation is insufficient, which is recognized by the RL optimization and leads to an increased mean value of $\alpha$ and thus more dissipative FV used within the scheme.
Thus, the found $\alpha$ distributions again indicate that the optimal closure is indeed dependent on the discretization schemes and must be adjusted to it.

\section{Conclusions}
\label{sec:conclusion}
Implicit filtering induces computational subgrid stresses, which introduce discretization-dependent closure terms into the LES equations. The discretization operators can have inhomogeneous approximation properties that make those stresses a function of space. Nonlinear discretization operators additionally give rise to nonlinear subfilter terms beyond the Reynolds stresses; a fact that is intentionally exploited for implicit closure modeling strategies. For both implicit and explicit modeling approaches however, this additional dependence of the unclosed terms on the actual discretization makes a priori SGS model development and analysis intractable. This inconsistency is at the heart of the discrepancy between a priori and a posteriori closure model performance. An alternative method to overcome these difficulties is to rely on a combination of physics-based closure models and optimization strategies that can account for these uncertainties. An optimization method that observes and interacts with the discrete dynamical system evoked by a specific discretization in an a posteriori manner has the potential for finding discretization-consistent closure models. Reinforcement learning is one such strategy.

In this work, we have formulated the task of finding closure models for implicitly filtered LES by optimizing their parameters as an RL task. In all cases, the found models gave consistently accurate results for considerably different discretization schemes and modeling approaches, which demonstrates the flexibility of the approach. As an example for an explicit model, a cell-wise constant or cell-wise quadratic Smagorinsky constant was optimized. The analysis of the RL-Quad results indicate that the RL agent recognizes the dissipation footprint of the underlying discretization. This is interesting, but clearly needs further analysis. As a novel implicit closure strategy, a blended hybrid DG/FV scheme was investigated that was originally intended for shock capturing. The found blending policy gives very accurate results for the proposed test case and warrants further study into these types of schemes as implicit modeling alternatives for compressible flows.

Our results clearly indicate that the proposed RL-based optimization scheme is capable of finding discretization- and model-specific closures that have the potential of at least partially solving the dilemma of discretization-induced uncertainty of the filter form. We do not propose these models directly as new, ready to be used closure models. Instead, we have shown a possible and flexible way of developing such models with universal performance for a specific discretization.
In order to proceed along those lines, a variety of challenges have to tackled that occur in all modeling approaches and are not specific to ML-augmented ones and comprise most importantly their generalizability, interpretability and uncertainties.

In addition, it would be beneficial to reduce the intitial hurdle of having to train the model under considerable computational effort for each specific discretization, but also for different flow physics.
A possible approach could be to first train the model in an offline setting to reproduce a well-performing analytical model, which then acts as an improved initial guess for the policy in the succeeding RL training in order to decrease the training time.
Another possible route is to apply transfer learning for RL~\cite{zhu2023transfer,taylor2009transfer}, since transfer learning was already applied successfully for turbulence modeling in a supervised setting~\cite{subel2021data}.
Here, it is exploited that a model trained for a distinct case adapts faster to a new training environment than it can learn a good policy from scratch.
The more sophisticated approach of meta learning for RL improves on naive transfer learning by not only training the agent on its original task, but also its ability to adapt quickly and efficiently to new, similar environments \cite{gupta2018meta}.

Since our proposed RL method can by design incorporate any existing modeling ideas, investigating existing models and their combinations is also a natural first step.

\begin{acknowledgments}
  This article may be downloaded for personal use only. Any other use requires prior permission of the author and AIP Publishing. This article appeared in \cite{beck2023toward} and may be found at \url{https://doi.org/10.1063/5.0176223}.

  The research presented in this paper was funded by Deutsche Forschungsgemeinschaft (DFG, German Research Foundation) under Germany’s Excellence Strategy - EXC 2075 -- 390740016. The work was also supported by funding from the  European High Performance Computing Joint Undertaking (JU) and the BMBF under the grant agreement No 101093393.
  The authors gratefully acknowledge the support and the computing time on ``Hawk'' provided by the HLRS through the project ``hpcdg'' and the support by the Stuttgart Center for Simulation Science (SimTech).
  The authors would also like to thank Philipp Offenh\"auser from HPE for the ongoing development efforts for the Relexi framework and Patrick Kopper for the implementation of the forced HIT test case in FLEXI.
\end{acknowledgments}

\section*{Author Declarations}
The authors have no conflicts to disclose.

\section*{Data Availability Statement}
The data that support the findings of this study are available from the corresponding author upon request. The LES code framework is available as open source at \url{https://github.com/flexi-framework/flexi}, the RL framework at \url{https://github.com/flexi-framework/relexi}.

\appendix
\section{Inference on Deformed Meshes}
\label{app:deformed}

\begin{figure}
  \ifthenelse{\boolean{final}}{
    \includegraphics[width=0.73\linewidth]{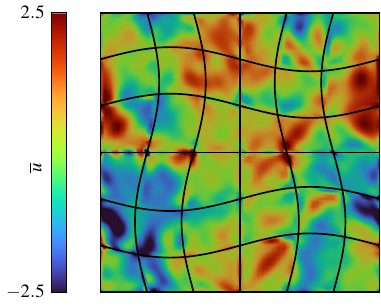}
  }{
    \tikzsetnextfilename{fig_tikz_deformed_mesh}
    \input{tikz/fig_deformed_mesh.tikz}
  }
  \caption{Slice of the interpolated intial solution on the deformed mesh at $x_1=3\pi/2$, where the deformation in $x_1$-direction is maximum. The mesh employs a high-order geometry representation with a polynomial degree of $N_{geo}=4$. The black lines illustrate the outlines of the mesh elements.}
  \label{fig:deformed_mesh}
\end{figure}

This section provides a brief discussion on how the proposed closure approaches translate to deformed meshes.
For this, the Cartesian mesh from the previous tests is deformed using the analytical function
\begin{equation}
  \ppvec{x}_d = \ppvec{x} + a\,\sin(x_1)\,\sin(x_2)\,\sin(x_3),
  \label{eq:deformed_mesh}
\end{equation}
where $\ppvec{x}=(x_1,x_2,x_3)$ denotes the nodes of the Cartesian mesh and $\ppvec{x}_d$ the nodes of its deformed version.
The amplitude of the deformation is set to $a=\pi/10$ and the period of the sinusoidal deformations is chosen as $2\pi$ such that the deformed mesh is also confined to the domain $\ppvec{x}_d\in[0,2\pi]^3$ and only the inner nodes are changed.
This allows to interpolate the flow fields easily from the undeformed to the deformed mesh and vice versa for the computation of the TKE spectra via a global Fourier transform.
The deformation is represented using a high-order geometry representation using a polynomial degree of $N_{geo}=4$.
The resulting deformed mesh for the case with 6 elements in each direction is shown in \figref{fig:deformed_mesh}.

To assess the performance of the explicit RL models discussed in \secref{sec:results}, LES computations with the different models are performed on the deformed mesh using the same initial state as in the Cartesian case.
It is important to stress that the RL-based models are not trained on or adapted to this new mesh geometry, but rather the models trained on the Cartesian mesh are applied as is on the deformed mesh.
The resulting TKE spectra are computed analogously to the procedure discussed in \secref{sec:results} and are given in \figref{fig:cs_comp_DOFs} with the SSM, DSM and NoMo closures shown for reference.
To compute the global energy spectrum, the solution was first interpolated from the deformed mesh to the Cartesian mesh using a high-order interpolation.
It seems important to note that this interpolation procedure between the undeformed and deformed meshes introduces errors since it requires to interpolate across the discontinuities between the individual DG elements, which are considerable in the underresolved LES case.
However, those errors seem negligible in the resulting flow field shown in \figref{fig:deformed_mesh} and also the derived quantities such as the TKE spectrum.

\begin{figure}
  \ifthenelse{\boolean{final}}{
    \includegraphics[width=0.95\linewidth]{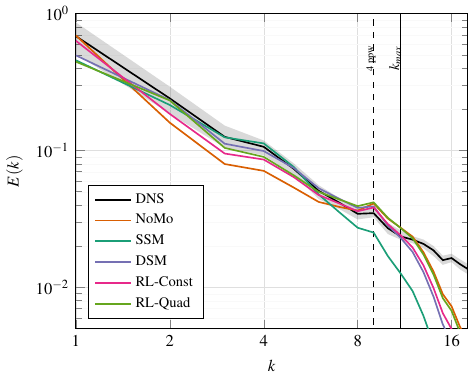}
  }{
    \tikzsetnextfilename{fig_tikz_spectra_deformed}
    \input{tikz/fig_spectra_deformed.tikz}
  }
  \caption{TKE spectra on the deformed mesh shown in \figref{fig:deformed_mesh} for the 36~DOF, $N=5$ case. The RL-Const and RL-Quad models werely solely trained on the non-deformed mesh as detailed in \secref{sec:results}. The results for the no-model LES, the SSM, the DSM on the deformed mesh and the DNS solution are shown for reference with the shaded area indicating the minimum and maximum observed TKE per wavenumber in the DNS.}
  \label{fig:app_deformed_results}
\end{figure}

The results in \figref{fig:app_deformed_results} indicate that all LES computations show reduced accuracy in comparison to the undeformed case.
This is to be expected since areas with low resolution capabilities are introduced by the stretching of elements which adds more numerical dissipation to the simulation.
This is especially evident for the low wavenumbers, where all simulations exhibit a reduction in TKE in comparison to the non-deformed case as shown in \figref{fig:cs_comp_DOFs}.
Despite this general trend, the qualitative behavior of the different models remains unchanged.
The no-model LES shows a significant drop in TKE for low wavenumbers and an accumulation of TKE near the cutoff wavenumber, while the SSM case shows the inverse behavior, where excessive dissipation in the high wavenumbers can be observed which leads to the accumulation of energy in the low and medium wavenumbers.
The DSM and the RL-Quad model yield comparable results, while the RL-Quad case exhibits slightly more TKE in the high wavenumbers near the cutoff. 
However, the DSM and RL-Quad models both outperform the RL-Const model, which shows a pronounced drop in TKE at medium wavenumbers around $k \in [2,6]$.
This is remarkable, since both the RL-Const and RL-Quad models show very similar results for the undeformed mesh as showcased in \figref{fig:cs_comp_DOFs}.
Since the deformed case is a more challenging test for the applied models, the obtained results might indicate that a non-uniform distribution of the model constant in the non-isotropic case becomes even more advantageous in contrast to the constant distribution.
However, these findings need to be confirmed by further investigations.

\bibliography{bibliography.bib}%

\end{document}